\documentclass[aps,twocolumn,showpacs,preprintnumbers,floats]{revtex4}\def\@cite#1#2{\textsuperscript{[{#1\if@tempswa , #2\fi}]}}
\usepackage{graphicx}
\usepackage{graphics}
\usepackage{amsmath}
\usepackage{amsfonts}
\usepackage{amssymb}
\usepackage{color}
\usepackage{subfigure}
\usepackage{epsfig}
\usepackage{morefloats}
\usepackage{multirow}
\usepackage{graphicx,booktabs}
\usepackage{mathrsfs}
\usepackage{txfonts}
\usepackage{indentfirst}
\usepackage{graphicx,booktabs}
\usepackage[colorlinks, citecolor=blue,anchorcolor=red,menucolor=red, linkcolor=red,filecolor=red,urlcolor=blue,frenchlinks=red]{hyperref}
\usepackage{longtable,lscape}
\usepackage{warpcol}
\usepackage{dcolumn}
\usepackage{longtable}

\begin{document}

\title{Mass spectrum and strong decays of strangeonium in a constituent quark model}
\author{ Qi Li, Long-Cheng Gui~\footnote {Mail: guilongcheng@hunnu.edu.cn}, Ming-Sheng Liu, Qi-Fang L\"{u}, Xian-Hui Zhong \footnote {Mail: zhongxh@hunnu.edu.cn}
 }  \affiliation{ 1) Department of Physics, Hunan Normal University,  Changsha 410081, China }

\affiliation{ 2) Synergetic Innovation
Center for Quantum Effects and Applications (SICQEA), Changsha 410081,China}

\affiliation{ 3) Key Laboratory of
Low-Dimensional Quantum Structures and Quantum Control of Ministry
of Education, Changsha 410081, China}

%\date{\today}

\begin{abstract}

In this work we calculate the mass spectrum of strangeonium up to the $3D$ multiplet within a nonrelativistic linear potential quark model. Furthermore, using the obtained wave functions, we also evaluate the strong decays of the strangeonium states with the $^3P_0$ model.
Based on our successful explanations of the well established states $\phi(1020)$, $\phi(1680)$, $h_1(1415)$,
$f'_2(1525)$, and $\phi_3(1850)$, we further discuss the possible assignments of
strangeonium-like states from experiments by combining our theoretical results
with the observations. It is found that some resonances, such as $f_2(2010)$ and $f_2(2150)$ listed by the Particle Data Group,
and $X(2062)$ and $X(2500)$ newly observed by BESIII, may be interpreted as the strangeonium states.
The possibility of $\phi(2170)$ as a candidate for $\phi(3S)$ or $\phi(2D)$ cannot be excluded.
We expect our results to provide useful references for looking for the missing $s\bar{s}$ states in future experiments.

\end{abstract}

%\begin{keyword}
%Spectrum, Strangeonium, Strong decays
%\end{keyword}

\maketitle

\section{Introduction}

The strangeonium ($s\bar{s}$) states, as one kind of quarkonium states predicted in the quark model,
lie between the light $q\bar{q}$ states and heavy charmonium ($c\bar{c}$) states. The $s\bar{s}$ states provide
a bridge for systematically exploring the Quantum Chromo-dynamics for light and heavy quarks.
Furthermore, the study of $s\bar{s}$ states is associated with the relative
topic of non-$q\bar{q}$ states (glueballs, hybrids, and tetraquarks etc.) with
the same quantum numbers as conventional $q\bar{q}$ system~\cite{Tanabashi:2018oca}.
To confirm a non-$q\bar{q}$ state from experiments,
one needs good knowledge of the conventional $q\bar{q}$ states.
However, at present data for the $s\bar{s}$ spectrum are rather scarce~\cite{Tanabashi:2018oca,Liu:2015zqa}. There are only
a few well established resonances $\phi(1020)$, $\phi(1680)$, $h_1(1415)$, $f_2'(1525)$ and $\phi_{3}(1850)$
from experiments which are widely accepted as the $s\bar{s}$ states. Besides some low-lying $1P$- and
$1D$-wave states, many $s\bar{s}$ states predicted in the quark model are waiting to be established.
For a long time the information of the $s\bar{s}$ states
was mainly extracted from the $\gamma p$, $K^-p$, $\pi^-p$, and $e^+e^-$ reactions.
The lack of data may be due to the fact that these experiments do not
efficiently produce $s\bar{s}$ states.

The BESIII experiments provide a powerful platform for
us to study the $s\bar{s}$ states. The world's largest $J/\psi$ and $\psi(2S)$
samples are best suited to study the $s\bar{s}$ spectrum via their decays~\cite{Liu:2015zqa,Yuan:2019zfo,Ablikim:2019hff}.
Recently, the BESIII Collaboration not only confirmed many
$s\bar{s}$ candidates observed in previous experiments, but also found
some new $s\bar{s}$ candidates by the decays of $J/\psi$ and $\psi(2S)$. For example, in 2019, evidence of a new $1^{+}$ resonance $X(2060)$ with
a mass of $M=(2062.8\pm13.1\pm7.2)$ MeV [or $1^{-}$ resonance $X(2000)$ with $M=(2002.1\pm27.5\pm15.0)$ MeV] was observed in $J/\psi\to \phi \eta \eta'$ at BESIII~\cite{Ablikim:2018xuz}. This resonance may be a candidate for the $2^1P_1$ (or $3^3S_1$~\cite{Pang:2019ttv}) $s\bar{s}$ state.
In 2018, by an amplitude analysis of the process $J/\psi\to \gamma K_SK_S$ several
isoscalar $0^{++}$ and $2^{++}$ states around $1.3-2.5$ GeV were extracted with a high significance
by the BEIII Collaboration, one broad $0^{++}$ state with a mass of $M=(2411\pm17)$ MeV and another broad $2^{++}$
state with a mass of $M=(2233\pm 34$$^{+9}_{-25})$ MeV might be candidates for the $3^3P_0$ and
$1^3F_2$ $s\bar{s}$ states, respectively~\cite{Ablikim:2018izx}. In 2016, several isoscalar $0^{-+}$,
$0^{++}$ and $2^{++}$ states around $2.0-2.4$ GeV were observed in $J/\psi\to \gamma \phi\phi$ at BESIII~\cite{Ablikim:2016hlu},
the $f_2(2010)$ confirmed in this process might be a candidate
for the $2^3P_2$ $s\bar{s}$ state; while the newly observed resonance $X(2500)$ might be a
candidate for a higher $0^{-+}$ $s\bar{s}$ state~\cite{Pan:2016bac} .
The world's most precise resonance parameters of $h_1(1415)$ were also determined by a recent measurement
of $J/\psi \to \eta' K\bar{K}\pi$ at BESIII~\cite{Ablikim:2018ctf}. Recently,
the vector meson resonance $\phi(2170)$ (often denoted as $Y (2175)$ in the literature)
was also confirmed in the $K^+(1460)K^-$, $K^+_1(1400)K^-$, $K^+_1(1270)K^-$, and
$\phi\eta'$ final states by the BESIII Collaboration~\cite{Ablikim:2020pgw,1788734},
this state might be a candidate for the $3^3S_1$ or $2^3D_1$ $s\bar{s}$ state
~\cite{Barnes:2002mu,Pang:2019ttv,Ding:2007pc,meson4,Coito:2009na}.
It should be mentioned that some forthcoming experiments from other collaborations such as COMPASS,
BelleII, GlueX, PANDA, and etc., will also provide more opportunities for us to study the $s\bar{s}$ states.

In theory, the $s\bar{s}$ mass spectrum was widely discussed within various quark models, for example,
the relativized quark model~\cite{Godfrey:1985xj,Xiao:2019qhl},
the nonrelativistic covariant oscillator quark model~\cite{Ishida:1986vn},
the QCD-motivated relativistic quark model~\cite{Ebert:2009ub},
the nonrelativistic constituent quark model constrained
in the study of the $NN$ phenomenology and the baryon spectrum~\cite{Vijande:2004he},
the nonrelativistic constituent quark potential model~\cite{Burakovsky:1997ci},
the extended Nambu-Jona-Lasinio quark model~\cite{Chizhov:2003qy,Chizhov:2020wug},
the approach of Regge trajectories~\cite{Anisovich:2000kxa,Badalian:2019xir},
the modified relativized quark model~\cite{Pang:2019ttv}, the framework of
the Bethe-Salpeter equation~\cite{Ricken:2003ua,Munz:1993si,Ricken:2000kf}, and so on.
Furthermore, the strong decay properties of the strangeonia
were studied within the pseudoscalar emission model~\cite{Godfrey:1985xj},
the flux-tube breaking model~\cite{Kokoski:1985is,Kumano:1988ga},
the $^3P_0$ model~\cite{Barnes:2002mu}, the corrected $^3P_0$ model~\cite{deQuadros:2020ntn},
the framework of relativistic quark model~\cite{Ricken:2003ua}, and so on.
However, a systematic study of both the $s\bar{s}$ mass spectrum and their decays by combining the
recent experimental progress is not found in the literature. An early review of
the status of the $s\bar{s}$ spectrum can be found in Ref.~\cite{Godfrey:1998pd}.

Stimulated by recent notable progress in experiments, we carry out a systematical
study of both the mass spectrum and strong decay properties of the $s\bar{s}$ system.
First, we calculate the mass spectrum up to the mass region of $3D$-wave states within a nonrelativistic
constituent quark potential model by partially adopting the model
parameters determined by the $\Omega$ spectrum~\cite{Liu:2019wdr}.
As done in the literature, e.g.~\cite{Deng:2016ktl,Li:2019tbn,Deng:2016stx},
the spin-dependent potentials are dealt with non-perturbatively so that
the effects of the spin-dependent interactions on the wave-functions can be included.
More importantly, with the widely used $^3P_0$ model~\cite{Micu:1968mk,LeYaouanc:1972vsx,LeYaouanc:1973ldf} we further analyze the Okubo-Zweig-lizuka (OZI)-allowed two-body strong decays of the $s\bar{s}$ states by using wave functions obtained from the potential model, which are crucial
to identify the nature of the resonances observed in experiments.
We obtain successful explanations of both the mass and strong decay properties for the well established states $\phi(1020)$, $\phi(1680)$, $h_1(1415)$, $f'_2(1525)$, and $\phi_3(1850)$. We find that (i) the $f_2(2010)$ and $f_2(2150)$ listed by the Particle Data Group (PDG)~\cite{Tanabashi:2018oca} might be candidates for the $2^3P_2$ and $1^3F_2$ $s\bar{s}$ states, respectively; (ii) the $4^{++}$ resonance $f_4(2210)$ first observed in the reaction $K^-p\to K^+K^- \Lambda$ by the LASS Collaboration~\cite{Aston:1988yp} might be
an assignment of the $1^3F_4$ $s\bar{s}$ state; (iii) the $f_0(2410)$ observed in $J/\psi\to K_SK_S$ at
BESIII~\cite{Ablikim:2018izx} may favor the assignment of the $3^3P_0$ $s\bar{s}$ state;
(iv) the newly observed resonances $X(2500)$~\cite{Ablikim:2016hlu} and $X(2062)$~\cite{Ablikim:2018xuz} from BESIII may be identified as the $4^1S_0$ and $2^1P_1$ $s\bar{s}$ states, respectively; (v) the possibility of $\phi(2170)$ as a candidate for $\phi(3S)$ or $\phi(2D)$ cannot be excluded, as the strong decay properties are very sensitive to its mass.

This paper is organized as follows. In Sec.~\ref{spectrum}, the mass spectrum is calculated within
a nonrelativistic linear potential model. Then, by using the obtained spectrum the OZI-allowed two-body strong decays
of the $s\bar{s}$ states are estimated in Sec.~\ref{Strongdecay} within the $^3P_0$ model.
In Sec.~\ref{DIS}, we discuss the properties of the $s\bar{s}$ states by combining our predictions
with the experimental observations or other model predictions. Finally, a summary is given in Sec.~\ref{sum}.

\begin{table}[htb]
\begin{center}
\caption{The parameters of the nonrelativistic potential model.}\label{parameter}
\begin{tabular}{ccccccccccccccccc}
\midrule[1.0pt]\midrule[1.0pt]
     & ~~~~~~~This work~~~~~~~ & ~~~~~~~Ref.~\cite{Liu:2019wdr}~~~~~~~   \\
\midrule[1.0pt]
   $m_s$ (GeV)        &   0.600        & Same   \\
   $\alpha_s$         &   0.770        & Same   \\
   $\sigma$ (GeV)     &   0.600        & Same   \\
   $b$      (GeV$^2$) &   0.135        & 0.110   \\
   $C_0$    (GeV)     &   $-0.519$     & $-0.694$    \\
\midrule[1.0pt]\midrule[1.0pt]
\end{tabular}
\end{center}
\end{table}

\section{mass spectrum}\label{spectrum}

To calculate the $s\bar{s}$ mass spectrum, we adopt a nonrelativistic potential model ~\cite{Barnes:2005pb,Eichten:2007qx,Li:2019tbn,Deng:2016stx,Deng:2016ktl}.
In this model, the effective quark-antiquark potential
is written as the sum of the spin-independent term $H_0(r)$ and spin-dependent term $H_{sd}(r)$; i.e.,
\begin{eqnarray}\label{H1}
V(r)=H_0(r)+H_{sd}(r),
\end{eqnarray}
where
\begin{eqnarray}\label{H0}
H_0(r)=-\frac{4}{3}\frac{\alpha_s}{r}+br+C_{0}
\end{eqnarray}
includes the standard color Coulomb interaction and the linear confinement.
The spin-dependent part $H_{sd}(r)$ can be expressed as~\cite{Eichten:1980mw}
\begin{eqnarray}\label{H0}
H_{sd}(r)=H_{SS}+H_{T}+H_{LS},
\end{eqnarray}
where
\begin{eqnarray}\label{ss}
H_{SS}= \frac{32\pi\alpha_s}{9m_qm_{\bar{q}}}\tilde{\delta}_\sigma(r)\mathbf{S}_{q}\cdot \mathbf{S}_{\bar{q}}
\end{eqnarray}
is the spin-spin contact hyperfine potential. Here, we take $\tilde{\delta}_\sigma(r)=(\sigma/\sqrt{\pi})^3
e^{-\sigma^2r^2}$ as suggested in Ref.~\cite{Barnes:2005pb}. The tensor potential $H_T$ is adopted as
\begin{eqnarray}\label{t}
H_{T}= \frac{4}{3}\frac{\alpha_s}{m_qm_{\bar{q}}}\frac{1}{r^3}\left(\frac{3\mathbf{S}_{q}\cdot \mathbf{r}\mathbf{S}_{\bar{q}}\cdot \mathbf{r}}{r^2}-\mathbf{S}_{q}\cdot\mathbf{S}_{\bar{q}}\right).
\end{eqnarray}
For convenience in the calculations, the potential of the spin-orbit interaction $H_{LS}$ is decomposed
into symmetric part $H_{sym}$ and antisymmetric part
$H_{anti}$,
\begin{eqnarray}\label{vs}
H_{LS}=H_{sym}+H_{anti},
\end{eqnarray}
with
\begin{eqnarray}\label{vs}
H_{sym}=\frac{\mathbf{S_{+}\cdot L}}{2}\left[\left(\frac{1}{2m_{\bar{q}}^{2}}+\frac{1}{2m_{q}^{2}}\right)\left(\frac{4}{3}\frac{\alpha_{s}}{r^{3}}-\frac{b}{r}\right)
+\frac{8\alpha_{s}}{3m_{q}m_{\bar{q}}r^{3}}\right],\\
H_{anti}=\frac{\mathbf{S_{-}\cdot L}}{2}\left(\frac{1}{2m_{q}^{2}}-\frac{1}{2m_{\bar{q}}^{2}}\right)\left(\frac{4}{3}\frac{\alpha_{s}}{r^{3}}-\frac{b}{r}\right).\ \ \ \ \ \ \ \ \ \ \ \ \ \ \ \ \ \ \ \ \ \ \
\end{eqnarray}
In these equations, $\mathbf{L}$ is the relative orbital angular momentum of the $q\bar{q}$
system; $\mathbf{S}_q$ and $\mathbf{S}_{\bar{q}}$ are the spins of the quark $q$ and antiquark $\bar{q}$, respectively, and $\mathbf{S}_{\pm}\equiv\mathbf{S}_q\pm \mathbf{S}_{\bar{q}}$; $m_q$ and $m_{\bar{q}}$ are the
masses of quark $q$ and antiquark $\bar{q}$, respectively; $\alpha_s$ is
the running coupling constant of QCD; and $r$ is the distance between the quark $q$ and antiquark $\bar{q}$, the constant $C_0$ stand for the zero point energy. The six parameters in the above equations ($\alpha_s$, $b$, $\sigma$, $m_q$, $m_{\bar{q}}$ $C_0$) are determined by fitting the spectrum.

Recently, the nonrelativistic potential model has been applied to study the $\Omega$ baryon spectrum~\cite{Liu:2019wdr}.
In order to be consistent with the $\Omega$ spectrum, we set the parameters $\alpha_s$, $\sigma$ and $m_s$
to the same values as those determinations in Ref.~\cite{Liu:2019wdr}.
The studies in Refs.~\cite{Godfrey:1985xj,Capstick:1986bm} show that the parameters $b$ and $C_0$ for the $q\bar{q}$ system might be slightly
different from the $qqq$ system, thus, in present work we reasonably adjust $b$ and $C_0$ to better describe
the masses of $s\bar{s}$ states $\phi(1020)$, $\phi(1680)$ and $\phi_{3}(1850)$.
Our parameters are listed in Table~\ref{parameter} where they are compared
to those of $\Omega$ baryon spectrum.

We solve the Schr\"{o}dinger equation by using
the three-point difference central method~\cite{Haicai} from central ($r=0$)
towards outside ($r\to \infty$) point by point. This method is successfully applied to the $b\bar{b}$, $\bar{b}c$ and $c\bar{c}$ systems~\cite{Deng:2016ktl,Deng:2016stx,Li:2019tbn}.
To overcome the singular behavior of $1/r^3$ in the
spin-dependent potentials, we introduce a cutoff distance $r_{c}$ in
the calculation. In a small range $r\in(0,r_{c})$, we let
$1/r^3=1/r_c^3$. With this treatment, one can deal
with spin-dependent potentials nonperturbatively so that the effects of the spin-dependent potentials on the wave-functions can be included.
Considering the fact that the mass of $1^3D_1$ is sensitive to the cutoff distance
$r_c$, the mass of $1^3D_1$ is used to determine the value of $r_c$. It should be
pointed out the $1^3D_1$ state is still not established in experiments.
Thus, we adopt a theoretical mass of $1^3D_1$ predicted with a method of perturbation, i.e., we let $H=H_{0}+H'$, where $H'$ is a part which contains the term of $1/r^{3}$. With the method of perturbation one can obtain a fairly accurate mass
although one cannot include the effects of the spin-dependent interactions on the wave-functions.
By solving the equation of $H_{0}|\psi_{n}^{(0)}\rangle=E_{0}|\psi_{n}^{(0)}\rangle$, we get the energy $E_{0}$ and wave function $|\psi_{n}^{(0)}\rangle$. Then, the mass of $1^3D_1$, 1809~MeV, is work out with $M=2m_s+E_{0}+\langle\psi_{n}^{(0)}|H'|\psi_{n}^{(0)}\rangle$.
Finally, with this predicted mass the cutoff distance $r_c$ is determined to be $0.546$ fm.

\begin{figure*}[!htbp]
\begin{center}
\centering  \epsfxsize=18.6cm \epsfbox{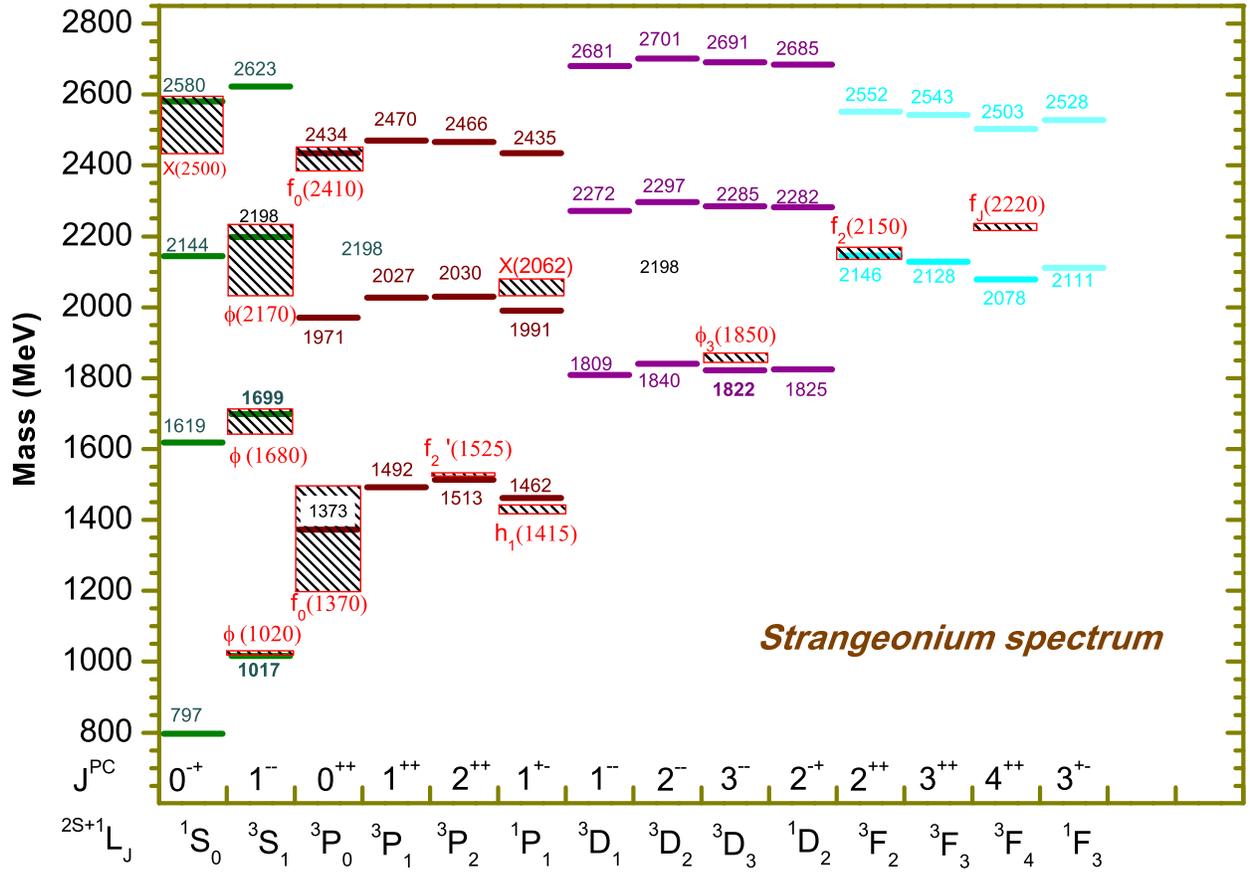}
\vspace{-1.8cm}\caption{The strangeonium spectrum predicted within a nonrelativistic linear potential quark model.
The shaded areas correspond to the experimental masses and their uncertainties, which are taken from the Particle Data Group~\protect\cite{Tanabashi:2018oca} and BESIII Collaboration~\protect\cite{Ablikim:2016hlu,Ablikim:2018xuz}. } \label{masses}
\end{center}
\end{figure*}

Our predicted $s\bar{s}$ mass spectrum is shown Fig~\ref{masses}. For a comparison, our results together with
some other model predictions and measurements are listed in Table~\ref{mass} as well.
From the table, one can see that the our predictions with
the nonrelativistic potential model are in reasonable agreement with other predictions of
the relativized quark model~\cite{Godfrey:1985xj,Xiao:2019qhl,Pang:2019ttv}, relativistic quark models~\cite{Ebert:2009ub}, and
nonrelativistic quark models~\cite{Ishida:1986vn,Vijande:2004he},
although some model dependencies exist in the predictions for the higher excitations with $n\geq 3$.
To understand why acceptable results can be provided by relativistic
as well as by nonrelativistic approaches, some people studied the connections
existing between relativistic, semirelativistic, and nonrelativistic
potential models of quarkonium using an interaction composed of an attractive Coulomb
potential and a confining power-law term~\cite{Semay:1992xq}.

\begin{figure}[!htbp]
\begin{center}
\centering  \epsfxsize=8.5cm \epsfbox{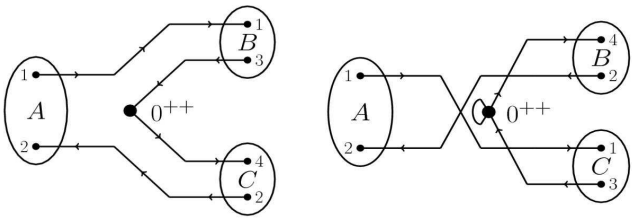}
\vspace{0cm}\caption{The meson two-body strong decay process $A \to BC$ in the $^3P_0$ model.} \label{3P0}
\end{center}
\end{figure}

\begin{table*}[htb]
\begin{center}
\caption{Predicted masses of the $s\bar{s}$ states in this work compared with the other predictions and observations. }\label{mass}

\begin{tabular}{ccccccccccccccccc}\midrule[1.0pt]\midrule[1.0pt]
~~~~~$n^{2S+1}L_{J}$~~~~~&$J^{PC}$~~~~~&Ours~~~~~& XWZZ~\cite{Xiao:2019qhl}~~~~~& EFG~\cite{Ebert:2009ub} ~~~~~& SIKY~\cite{Ishida:1986vn}&
~~~~~GI~\cite{Godfrey:1985xj}~~~~~&Pang~\cite{Pang:2019ttv}~~~~~&VFV~\cite{Vijande:2004he}~~~~~&Observed state ~\cite{Tanabashi:2018oca}\\
\midrule[1.0pt]\ %             me        xiao        ebert        ISKY           GI         pang	       exp
$1 ^3S_{1}$~~~~~&$1^{--}$ ~~~~~&1017 ~~~~~&1009   ~~~~~&1038    ~~~~~&1020    ~~~~~ &1020     ~~~~~&1030    ~~~~~&1020    ~~~~~&$\phi(1020)$        \\
$1 ^1S_{0}$~~~~~&$0^{-+}$ ~~~~~&797  ~~~~~&657    ~~~~~&743     ~~~~~&690     ~~~~~ &960      ~~~~~&$\cdots$~~~~~&956     ~~~~~&$\cdots$    \\
$2 ^3S_{1}$~~~~~&$1^{--}$ ~~~~~&1699 ~~~~~&1688   ~~~~~&1698    ~~~~~&1740    ~~~~~ &1690     ~~~~~&1687    ~~~~~&1726    ~~~~~&$\phi(1680)$ \\
$2 ^1S_{0}$~~~~~&$0^{-+}$ ~~~~~&1619 ~~~~~&1578   ~~~~~&1536    ~~~~~&1440    ~~~~~ &1630     ~~~~~&$\cdots$~~~~~&1795    ~~~~~&$\cdots$    \\
$3 ^3S_{1}$~~~~~&$1^{--}$ ~~~~~&2198 ~~~~~&2204   ~~~~~&2119    ~~~~~&2250    ~~~~~ &$\cdots$ ~~~~~&2149    ~~~~~&$\cdots$~~~~~&$\cdots$    \\
$3 ^1S_{0}$~~~~~&$0^{-+}$ ~~~~~&2144 ~~~~~&2125   ~~~~~&2085    ~~~~~&1970    ~~~~~ &$\cdots$ ~~~~~&$\cdots$~~~~~&$\cdots$~~~~~&$\cdots$    \\
$4 ^3S_{1}$~~~~~&$1^{--}$ ~~~~~&2623 ~~~~~&2627   ~~~~~&2472    ~~~~~&2540    ~~~~~ &$\cdots$ ~~~~~&2498    ~~~~~&$\cdots$~~~~~&$\cdots$    \\
$4 ^1S_{0}$~~~~~&$0^{-+}$ ~~~~~&2580 ~~~~~&2568   ~~~~~&2439    ~~~~~&2260    ~~~~~ &$\cdots$ ~~~~~&$\cdots$~~~~~&$\cdots$~~~~~& $X(2500)$~\cite{Ablikim:2016hlu}    \\
$1 ^3P_{2}$~~~~~&$2^{++}$ ~~~~~&1513 ~~~~~&1539   ~~~~~&1529    ~~~~~&1480    ~~~~~ &1530     ~~~~~&$\cdots$~~~~~&1556    ~~~~~&$f_2'(1525)$  \\
$1 ^3P_{1}$~~~~~&$1^{++}$ ~~~~~&1492 ~~~~~&1480   ~~~~~&1464    ~~~~~&1430    ~~~~~ &1480     ~~~~~&$\cdots$~~~~~&1508    ~~~~~& $f_1(1420)?$       \\
$1 ^3P_{0}$~~~~~&$0^{++}$ ~~~~~&1373 ~~~~~&1355   ~~~~~&1420    ~~~~~&1180    ~~~~~ &1360     ~~~~~&$\cdots$~~~~~&$\cdots$~~~~~&$f_0(1370)$ \\
$1 ^1P_{1}$~~~~~&$1^{+-}$ ~~~~~&1462 ~~~~~&1473   ~~~~~&1485    ~~~~~&1460    ~~~~~ &1470     ~~~~~&$\cdots$~~~~~&1511    ~~~~~&$h_1(1415)$  \\
$2 ^3P_{2}$~~~~~&$2^{++}$ ~~~~~&2030 ~~~~~&2046   ~~~~~&2030    ~~~~~&2080    ~~~~~ &2040     ~~~~~&$\cdots$~~~~~&1999    ~~~~~&$f_2(2010)$    \\
$2 ^3P_{1}$~~~~~&$1^{++}$ ~~~~~&2027 ~~~~~&2027   ~~~~~&2016    ~~~~~&2020    ~~~~~ &2030     ~~~~~&$\cdots$~~~~~&$\cdots$~~~~~&$\cdots$    \\
$2 ^3P_{0}$~~~~~&$0^{++}$ ~~~~~&1971 ~~~~~&1986   ~~~~~&1969    ~~~~~&1800    ~~~~~ &1990     ~~~~~&$\cdots$~~~~~&$\cdots$~~~~~&$\cdots$    \\
$2 ^1P_{1}$~~~~~&$1^{+-}$ ~~~~~&1991 ~~~~~&2008   ~~~~~&2024    ~~~~~&2040    ~~~~~ &2010     ~~~~~&$\cdots$~~~~~&1973    ~~~~~&$X(2062)$~\cite{Ablikim:2018xuz}    \\
$3 ^3P_{2}$~~~~~&$2^{++}$ ~~~~~&2466 ~~~~~&2480   ~~~~~&2412    ~~~~~&2540    ~~~~~ &$\cdots$ ~~~~~&$\cdots$~~~~~&$\cdots$~~~~~&$\cdots$    \\
$3 ^3P_{1}$~~~~~&$1^{++}$ ~~~~~&2470 ~~~~~&2468   ~~~~~&2403    ~~~~~&2480    ~~~~~ &$\cdots$ ~~~~~&$\cdots$~~~~~&$\cdots$~~~~~&$\cdots$    \\
$3 ^3P_{0}$~~~~~&$0^{++}$ ~~~~~&2434 ~~~~~&2444   ~~~~~&2364    ~~~~~&2280    ~~~~~ &$\cdots$ ~~~~~&$\cdots$~~~~~&$\cdots$~~~~~&$f_0(2410)$~\cite{Ablikim:2018izx}    \\
$3 ^1P_{1}$~~~~~&$1^{+-}$ ~~~~~&2435 ~~~~~&2449   ~~~~~&2398    ~~~~~&2490    ~~~~~ &$\cdots$ ~~~~~&$\cdots$~~~~~&$\cdots$~~~~~&$\cdots$    \\
$1 ^3D_{3}$~~~~~&$3^{--}$ ~~~~~&1822 ~~~~~&1897   ~~~~~&1950    ~~~~~&1830    ~~~~~ &1900     ~~~~~&$\cdots$~~~~~&1875    ~~~~~&$\phi_3(1850)$  \\
$1 ^3D_{2}$~~~~~&$2^{--}$ ~~~~~&1840 ~~~~~&1904   ~~~~~&1908    ~~~~~&1810    ~~~~~ &1910     ~~~~~&$\cdots$~~~~~&$\cdots$~~~~~&$\cdots$    \\
$1 ^3D_{1}$~~~~~&$1^{--}$ ~~~~~&1809 ~~~~~&1883   ~~~~~&1845    ~~~~~&1750    ~~~~~ &1880     ~~~~~&1869    ~~~~~&$\cdots$~~~~~&$\cdots$    \\
$1 ^1D_{2}$~~~~~&$2^{-+}$ ~~~~~&1825 ~~~~~&1893   ~~~~~&1909    ~~~~~&1830    ~~~~~ &1890     ~~~~~&$\cdots$~~~~~&1853    ~~~~~&$\cdots$    \\
$2 ^3D_{3}$~~~~~&$3^{--}$ ~~~~~&2285 ~~~~~&2337   ~~~~~&2338    ~~~~~&2360    ~~~~~ &$\cdots$ ~~~~~&$\cdots$~~~~~&$\cdots$~~~~~&$\cdots$    \\
$2 ^3D_{2}$~~~~~&$2^{--}$ ~~~~~&2297 ~~~~~&2348   ~~~~~&2323    ~~~~~&2330    ~~~~~ &$\cdots$ ~~~~~&$\cdots$~~~~~&$\cdots$~~~~~&$\cdots$    \\
$2 ^3D_{1}$~~~~~&$1^{--}$ ~~~~~&2272 ~~~~~&2342   ~~~~~&2258    ~~~~~&2260    ~~~~~ &$\cdots$ ~~~~~&2276    ~~~~~&$\cdots$~~~~~&$\cdots$    \\
$2 ^1D_{2}$~~~~~&$2^{-+}$ ~~~~~&2282 ~~~~~&2336   ~~~~~&2321    ~~~~~&2340    ~~~~~ &$\cdots$ ~~~~~&$\cdots$~~~~~&$\cdots$~~~~~&$\cdots$    \\
$3 ^3D_{3}$~~~~~&$3^{--}$ ~~~~~&2691 ~~~~~&2725   ~~~~~&2727    ~~~~~&$\cdots$~~~~~ &$\cdots$ ~~~~~&$\cdots$~~~~~&$\cdots$~~~~~&$\cdots$    \\
$3 ^3D_{2}$~~~~~&$2^{--}$ ~~~~~&2701 ~~~~~&2734   ~~~~~&2667    ~~~~~&$\cdots$~~~~~ &$\cdots$ ~~~~~&$\cdots$~~~~~&$\cdots$~~~~~&$\cdots$    \\
$3 ^3D_{1}$~~~~~&$1^{--}$ ~~~~~&2681 ~~~~~&2732   ~~~~~&2607    ~~~~~&$\cdots$~~~~~ &$\cdots$ ~~~~~&2593    ~~~~~&$\cdots$~~~~~&$\cdots$    \\
$3 ^1D_{2}$~~~~~&$2^{-+}$ ~~~~~&2685 ~~~~~&2723   ~~~~~&2662    ~~~~~&$\cdots$~~~~~ &$\cdots$ ~~~~~&$\cdots$~~~~~&$\cdots$~~~~~&$\cdots$    \\
$1 ^3F_{4}$~~~~~&$4^{++}$ ~~~~~&2078 ~~~~~&2202   ~~~~~&2286    ~~~~~&2130    ~~~~~ &2200     ~~~~~&$\cdots$~~~~~&$\cdots$~~~~~&$f_4(2210)$~\cite{Aston:1988yp}    \\
$1 ^3F_{3}$~~~~~&$3^{++}$ ~~~~~&2128 ~~~~~&2234   ~~~~~&2215    ~~~~~&2120    ~~~~~ &2230     ~~~~~&$\cdots$~~~~~&$\cdots$~~~~~&$\cdots$    \\
$1 ^3F_{2}$~~~~~&$2^{++}$ ~~~~~&2146 ~~~~~&2243   ~~~~~&2143    ~~~~~&2090    ~~~~~ &2240     ~~~~~&$\cdots$~~~~~&$\cdots$~~~~~&$f_2(2150)$    \\
$1 ^1F_{3}$~~~~~&$3^{+-}$ ~~~~~&2111 ~~~~~&2223   ~~~~~&2209    ~~~~~&2130    ~~~~~ &2220     ~~~~~&$\cdots$~~~~~&$\cdots$~~~~~&$\cdots$    \\
$2 ^3F_{4}$~~~~~&$4^{++}$ ~~~~~&2503 ~~~~~&2596   ~~~~~&2657    ~~~~~&$\cdots$~~~~~ &$\cdots$ ~~~~~&$\cdots$~~~~~&$\cdots$~~~~~&$\cdots$ \\
$2 ^3F_{3}$~~~~~&$3^{++}$ ~~~~~&2543 ~~~~~&2623   ~~~~~&2585    ~~~~~&$\cdots$~~~~~ &$\cdots$ ~~~~~&$\cdots$~~~~~&$\cdots$~~~~~&$\cdots$    \\
$2 ^3F_{2}$~~~~~&$2^{++}$ ~~~~~&2552 ~~~~~&2636   ~~~~~&2514    ~~~~~&$\cdots$~~~~~ &$\cdots$ ~~~~~&$\cdots$~~~~~&$\cdots$~~~~~&$\cdots$    \\
$2 ^1F_{3}$~~~~~&$3^{+-}$ ~~~~~&2528 ~~~~~&2613   ~~~~~&2577    ~~~~~&$\cdots$~~~~~ &$\cdots$ ~~~~~&$\cdots$~~~~~&$\cdots$~~~~~&$\cdots$    \\
\midrule[1.0pt]\midrule[1.0pt]
\end{tabular}
\end{center}
\end{table*}

\section{strong decays}\label{Strongdecay}
%\subsection{$^3P_0$ model}

In this work the Okubo-Zweig-lizuka (OZI) allowed two body strong decays of the $s\bar{s}$ states are calculated with  the widely used $^{3}P_0$ model~\cite{Micu:1968mk,LeYaouanc:1972vsx,LeYaouanc:1973ldf}. In this model, one assumes that a quark-antiquark pair is produced from
the vacuum with the quantum number $0^{++}$ and the initial meson decay takes place via the rearrangement of the four quarks as shown in Fig~\ref{3P0}. In the nonrelativistic limit, the transition operator is expressed as
\begin{eqnarray}
    \hat{T} & = & -3 \gamma \sqrt{96 \pi} \sum_{m}^{} \langle 1 m 1 -m| 0 0 \rangle \int_{}^{} d\mathbf{p}_3 d\mathbf{p}_4 \delta^3 (\mathbf{p}_3 + \mathbf{p}_4) \nonumber\\
      & \times &  \mathcal{Y}_1^m \left(\frac{\mathbf{p}_3 - \mathbf{p}_4}{2}\right)  \chi_{1-m}^{34}  \phi_0^{34} \omega_0^{34} b_{3i}^\dagger (\mathbf{p}_3) d_{4j}^\dagger (\mathbf{p}_4) \ ,
\end{eqnarray}
where $\gamma$ is a dimensionless constant that denotes the strength of the quark-antiquark pair creation with
momentum $\mathbf{p}_3$ and $\mathbf{p}_4$ from vacuum; $b_{3i}^\dagger (\mathbf{p}_3)$ and $d_{4j}^\dagger(\mathbf{p}_4)$ are the creation operators for the quark and antiquark, respectively; the subscripts, $i$ and $j$, are the SU(3)-color indices of the created quark and antiquark;
$\phi_0^{34}=(u\bar u +d\bar d +s \bar s)/\sqrt 3$ and $\omega_{0}^{34}=\frac{1}{\sqrt{3}} \delta_{ij}$ correspond to flavor and
color singlets, respectively; $\chi_{{1,-m}}^{34}$ is a spin triplet
state; and $\mathcal{Y}_{\ell m}(\mathbf{k})\equiv
|\mathbf{k}|^{\ell}Y_{\ell m}(\theta_{\mathbf{k}},\phi_{\mathbf{k}})$ is the
$\ell$-th solid harmonic polynomial.

For an OZI allowed two-body strong decay process $A\to B+C$, the helicity amplitude
$\mathcal{M}^{M_{J_A}M_{J_B} M_{J_C}}(\mathbf{P})$ can be worked out by
\begin{eqnarray}
\langle BC|T| A\rangle=\delta(\mathbf{P}_A-\mathbf{P}_B-\mathbf{P}_C)\mathcal{M}^{M_{J_A}M_{J_B} M_{J_C}}(\mathbf{P}).
\end{eqnarray}
In the center-of-mass (c.m.) frame of the initial meson $A$, the helicity amplitude can be written as
\begin{widetext}
\begin{eqnarray}\label{T-matrix}
\mathcal{M}^{M_{J_A}M_{J_B} M_{J_C}}(\mathbf{P})&=&\gamma \sqrt{96 \pi} \sum_{\begin{subarray}{l}
      M_{L_A},M_{S_A},
      M_{L_B},M_{S_B},\\
      M_{L_C},M_{S_C},m
      \end{subarray}}
\langle L_A M_{L_A}; S_A M_{S_A} | J_A M_{J_A} \rangle
\times \langle 1\;m;1\;-m|\;0\;0 \rangle\langle L_B M_{L_B};
S_B M_{S_B} | J_B M_{J_B} \rangle\nonumber\\
&&
\times\langle
L_C M_{L_C}; S_C M_{S_C} | J_C M_{J_C} \rangle\times
\langle \chi^{1 3}_{S_B M_{S_B}}\chi^{2 4}_{S_CM_{S_C}}  | \chi^{1 2}_{S_A M_{S_A}} \chi^{3 4}_{1 -\!m} \rangle\nonumber\\
&&\times[\langle\phi^{13}_B \phi^{2 4}_C | \phi^{1 2}_A \phi^{3 4}_0\rangle I^{M_{L_A},m}_{M_{L_B},M_{L_C}}(\mathbf{P})
+(-1)^{S_A+S_B+S_C+1}\langle\phi^{2 4}_B \phi^{1 3}_C | \phi^{1 2}_A \phi^{3 4}_0\rangle I^{M_{L_A},m}_{M_{L_B},M_{L_C}}(\mathbf{-P)}],
\end{eqnarray}
\end{widetext}
with the integral in the momentum space
\begin{eqnarray}\label{eq4}
I^{M_{L_A},m}_{M_{L_B},M_{L_C}}(\mathbf{P})=\int d^3\mathbf{p}_3 \Psi^*_{n_B L_B M_{L_B}}\left(\frac{m_3 \mathbf{P}}{m_1+m_3}-\mathbf{p}_3 \right)\mathcal{Y}_{1m}(\mathbf{p}_3)\nonumber\\
\times\Psi^*_{n_C L_C M_{L_C}}\left(\frac{-m_3 \mathbf{P}}{m_2+m_3}+\mathbf{p}_3 \right)\Psi_{n_A L_A M_{L_A}}\left(\mathbf{P}-\mathbf{p}_3 \right).
\end{eqnarray}
In the above equations, ($J_{A}$, $J_{B}$ and $J_{C}$), ($L_A$, $L_B$ and $L_C$) and ($S_A$, $S_B$ and $S_C$) are the quantum numbers of the total angular momentum, orbital angular momentum and total spin for hadrons $A,B,C$, respectively;
in the c.m. frame of hadron
$A$, the momenta $\mathbf{P}_B$ and $\mathbf{P}_C$ of mesons $B$ and $C$ satisfy
$\mathbf{P}_B=-\mathbf{P}_C\equiv \mathbf{P}$; $m_1$ and $m_2$ are the
constituent quark masses of the initial hadron $A$; $m_3$ is
the mass of the anti-quark created from vacuum;
$\Psi_{n_A L_A M_{L_A}}$, $\Psi_{n_B L_B M_{L_B}}$ and $\Psi_{n_C L_C M_{L_C}}$
are the radial wave functions of hadrons $A$, $B$ and $C$, respectively, in
the momentum space, while $\phi^{12}_A$, $\phi^{13}_B$ and $\phi^{24}_C$ ($\chi^{1 2}_{S_A M_{S_A}}$, $\chi^{1 3}_{S_B M_{S_B}}$
and $\chi^{2 4}_{S_C M_{S_C}}$) are the flavor (spin) wave functions
of hadrons $A$, $B$ and $C$, respectively;
$\langle \phi^{1
3}_B \phi^{2 4}_C | \phi^{1 2}_A \phi^{3 4}_{00}\rangle$ and $\langle  \chi^{1 3}_{S_B M_{S_B}} \chi^{2 4}_{S_C M_{S_C}}|
\chi^{1 2}_{S_A M_{S_A}} \chi^{3 4}_{1 -\!m} \rangle$ are the flavor and spin matrix elements, respectively;
$\langle L_A M_{L_A}; S_A M_{S_A} | J_A M_{J_A} \rangle$ and $\langle
L_B M_{L_B}; S_B M_{S_B} | J_B M_{J_B} \rangle$, $\langle
L_C M_{L_C}; S_C M_{S_C} | J_C M_{J_C} \rangle$ and
$\langle 1\;m;1\;-m|\;0\;0 \rangle$ are the corresponding Clebsch-Gordan
coefficients.

With the Jacob-Wick formula~\cite{Jacob:1959at},
the helicity amplitudes
$\mathcal{M}^{M_{J_A}M_{J_B} M_{J_C}}(\mathbf{P})$ can be converted to the partial wave amplitudes $\mathcal{M}^{JL}$ via
\begin{eqnarray}
{\mathcal{M}}^{J L}(A\rightarrow BC) = \frac{\sqrt{4\pi (2 L+1)}}{2 J_A+1}  \sum_{M_{J_B},M_{J_C}} \langle L 0 J M_{J_A}|J_A M_{J_A}\rangle\nonumber\\
\times  \langle J_B M_{J_B} J_C M_{J_C} | J M_{J_A} \rangle \mathcal{M}^{M_{J_A} M_{J_B} M_{J_C}}({\textbf{P}}),
\end{eqnarray}
where $M_{J_A}=M_{J_B}+M_{J_C}$ ,\;$\mathbf{J}\equiv \mathbf{J}_B+\mathbf{J}_C$ and
$\mathbf{J}_{A} \equiv \mathbf{J}_{B}+\mathbf{J}_C+\mathbf{L}$. More details of the $^3P_0$ model can be found
in our recent paper~\cite{Gui:2018rvv}.

To partly remedy the inadequacy of the nonrelativistic
wave function as the momentum $\mathbf{P}$ increases, the partial width of the $A\to B+C$ process is calculated with
a semirelativistic phase space~\cite{Kumano:1988ga,Kokoski:1985is}:
\begin{eqnarray}
\Gamma = 2\pi |\textbf{P}| \frac{M_B M_C}{M_A}\sum_{JL}\Big
|\mathcal{M}^{J L}\Big|^2,\label{de}
\end{eqnarray}
where $M_A$ is the mass of the initial hadron $A$, while $M_B$ and $M_C$ stand for the masses of
final hadrons $B$ and $C$, respectively. In our calculation, the masses of final hadrons
$B$ and $C$ appearing in the phase space are adopted the ``mock" values as suggested in Ref.~\cite{Kokoski:1985is}, they are worked out by
%\begin{widetext}
\begin{eqnarray}\label{Mock}
\tilde{M}(fnL)=\frac{1}{4}M(fnL;S=0,J=L)\ \ \ \ \ \ \ \ \ \ \ \ \ \ \ \ \ \ \ \ \ \ \ \ \ \ \ \ \nonumber\\
              +{\sum_{m=-1,0,1}\frac{2(L+m)+1}{4(2L+1)} M(fnL;S=1,J=L+m)},
\end{eqnarray}
%\end{widetext}
where $\tilde{M}$, $M$, $n$, $L$, $S$, $m$ stand for the mock mass, mass, principal quantum number, orbital quantum number,
spin quantum number, and the third component of spin momentum, respectively. The masses and ``mock" masses of final meson states have been collected in Table~\ref{Finalstates}.
Furthermore, the masses of the initial hadrons when known
experimentally are adopted from the PDG~\cite{Tanabashi:2018oca},
otherwise their masses are taken from our potential model predictions listed in Table~\ref{mass}.

In the calculations, the wave functions of the initial and final states are taken
from our potential model predictions. To obtain the
wave functions for the final kaon meson states, we fit the masses of the well established states $K$, $K^*$, $K(1460)$, $K_2(1430)$, $K^*(1680)$, and $K_3(1780)$. The potential model parameters for the kaon meson spectrum are determined to be $\alpha_s=0.885$, $b=0.1383$GeV$^2$, $m_s=0.6$ GeV,  $m_u=0.45$ GeV, $\sigma=0.669$ GeV, $r_c=0.59$ fm and $C_0=-0.524$ GeV.
The pair creation strength $\gamma=0.360$ is obtained by fitting the width of $\phi(1680)$
from the PDG~\cite{Tanabashi:2018oca}. The OZI-allowed two body strong decay properties,
such as the partial decay width, total decay width, branching fraction, and partial wave amplitude,
for the $s\bar{s}$ states listed in  Table~\ref{mass}
are calculated, our results have been listed in Tables~\ref{decay1}- \ref{decayfb}.

\begin{table*}[htb]
\begin{center}
\caption{Masses, mock masses (denoted with $\tilde{M}$ ) and flavor wave functions for the final meson states.
The meson masses for the well established states are adopted from the PDG~\cite{Tanabashi:2018oca}, otherwise they are taken from
our theoretical estimations. The mock masses, which correspond to the calculated mass of the meson in the spin-independent
$q\bar{q}$ potential, are calculated with Eq.~(\ref{Mock}). The mixing angle $\theta_{1P}=45^\circ$ and
$\theta_{1D}=45^\circ$ for the $1P$- and $1D$-wave kaon meson states are adopted the determinations according to the decay properties~\cite{Tanabashi:2018oca,Pang:2017dlw}.
}   \label{Finalstates}
\begin{tabular}{ccccccccccccccccc}\midrule[1.0pt]\midrule[1.0pt]
Meson ~~~&$n^{2S+1}L_J$    ~~~&$J^{P(C)}$    & Mass (MeV)    & $\tilde{M}$ (MeV)  ~~~& Flavor function  \\
\midrule[1.0pt]
$K$              ~~~&$1 ^1S_{0}$                                                                 ~~~&$0^{-}$     ~~~&494       ~~~& 793  ~~~&$K^+=u\bar{s},\,K^-=\bar{u}s,\,K^0=d\bar{s},\,\bar{K^0}=\bar{d}s$                                  \\
$K^*$            ~~~&$1 ^3S_{1}$                                                                 ~~~&$1^{-}$     ~~~&896       ~~~& 793  ~~~&$K^{*+}=u\bar{s},\,K^{*-}=\bar{u}s,\,K^{*0}=d\bar{s},\,\bar{K^{*0}}=\bar{d}s$                      \\
$K(1460)$        ~~~&$2 ^1S_{0}$                                                                 ~~~&$0^{-}$     ~~~&1460      ~~~& 1580 ~~~&$K^+=u\bar{s},\,K^-=\bar{u}s,\,K^0=d\bar{s},\,\bar{K^0}=\bar{d}s$                                  \\
$K^*(1410)$      ~~~&$2 ^3S_{1}$                                                                 ~~~&$1^{-}$     ~~~&1580      ~~~& 1580 ~~~&$K^{*+}=u\bar{s},\,K^{*-}=\bar{u}s,\,K^{*0}=d\bar{s},\,\bar{K^{*0}}=\bar{d}s$                      \\
$K^*_0(1430)$    ~~~&$1 ^3P_{0}$                                                                 ~~~&$0^{+}$     ~~~&1425      ~~~& 1381 ~~~&$K^{*+}_0=u\bar{s},\,K^{*-}_0=\bar{u}s,\,K^{*0}_0=d\bar{s},\,\bar{K^{*0}_0}=\bar{d}s$              \\
$K_1(1270)$      ~~~&$\cos\theta_{1P}|1\;^{1}P_1\rangle +\sin\theta_{1P}|1\;^{3}P_1\rangle$      ~~~&$1^{+}$     ~~~&1272      ~~~& 1381 ~~~&$K^+_1=u\bar{s},\,K^-_1=\bar{u}s,\,K^0_1=d\bar{s},\,\bar{K^0_1}=\bar{d}s$                          \\
$K_1(1400)$      ~~~&$ -\sin\theta_{1P}|1\;^{1}P_1\rangle +\cos\theta_{1P}|1\;^{3}P_1\rangle$    ~~~&$1^{+}$     ~~~&1403      ~~~& 1381 ~~~&$K^+_1=u\bar{s},\,K^-_1=\bar{u}s,\,K^0_1=d\bar{s},\,\bar{K^0_1}=\bar{d}s$                          \\
                 ~~~&$\theta_{1P}=45^\circ$~\cite{Tanabashi:2018oca,Pang:2017dlw}                             ~  &             ~ &           ~ &       ~ &                                                                                                   \\
$K^*_2(1430)$    ~~~&$1 ^3P_{2}$                                                                 ~~~&$2^{+}$     ~~~&1426      ~~~& 1381 ~~~&$K^{*+}_2=u\bar{s},\,K^{*-}_2=\bar{u}s,\,K^{*0}_2=d\bar{s},\,\bar{K^{*0}_2}=\bar{d}s$              \\
$K^*(1680)$      ~~~&$1 ^3D_{1}$                                                                 ~~~&$1^{-}$     ~~~&1718      ~~~& 1756 ~~~&$K^{*+}=u\bar{s},\,K^{*-}=\bar{u}s,\,K^{*0}=d\bar{s},\,\bar{K^{*0}}=\bar{d}s$                      \\
$K_2(1770)$      ~~~&$\cos\theta_{1D}|1\;^{1}D_2\rangle +\sin\theta_{1D}|1\;^{3}D_2\rangle$      ~~~&$2^{-}$     ~~~&1773      ~~~& 1756 ~~~&$K^+_2=u\bar{s},\,K^-_2=\bar{u}s,\,K^0_2=d\bar{s},\,\bar{K^0_2}=\bar{d}s$                          \\
$K_2(1820)$      ~~~&$ -\sin\theta_{1D}|1\;^{1}D_2\rangle +\cos\theta_{1D}|1\;^{3}D_2\rangle$    ~~~&$2^{-}$     ~~~&1819      ~~~& 1756 ~~~&$K^+_2=u\bar{s},\,K^-_2=\bar{u}s,\,K^0_2=d\bar{s},\,\bar{K^0_2}=\bar{d}s$                          \\
                 ~~~&$\theta_{1D}=45^\circ$~\cite{Pang:2017dlw}                             ~~~&            ~~~&          ~~~&      ~~~&                                                                                                   \\
$K^*_3(1780)$    ~~~&$1 ^3D_{3}$                                                                 ~~~&$3^{-}$     ~~~&1776      ~~~& 1756 ~~~&$K^{*+}_3=u\bar{s},\,K^{*-}_3=\bar{u}s,\,K^{*0}_3=d\bar{s},\,\bar{K^{*0}_3}=\bar{d}s$              \\
$\eta$           ~~~&$1 ^1S_{0}$                                                                 ~~~&$0^{-+}$    ~~~&548       ~~~& 793  ~~~&$\cos\theta_1(\frac{u\bar{u}+d\bar{d}}{\sqrt{2}})-\sin\theta_1({s\bar{s}}),\,\theta_1=39.3^\circ$  \\
$\eta'$          ~~~&$1 ^1S_{0}$                                                                 ~~~&$0^{-+}$    ~~~&958       ~~~& 793  ~~~&$\sin\theta_1(\frac{u\bar{u}+d\bar{d}}{\sqrt{2}})+\cos\theta_1({s\bar{s}}),\,\theta_1=39.3^\circ$  \\
$\phi(1020)$     ~~~&$1 ^3S_{1}$                                                                 ~~~&$1^{--}$    ~~~&1020      ~~~& 964  ~~~&$s\bar{s}$                                                                                         \\
$f_0(1373)$      ~~~&$1 ^3P_{0}$                                                                 ~~~&$0^{++}$    ~~~&1373      ~~~& 1488 ~~~&$s\bar{s}$                                                                                         \\
$f_1(1492)$      ~~~&$1 ^3P_{1}$                                                                 ~~~&$1^{++}$    ~~~&1492      ~~~& 1488 ~~~&$s\bar{s}$                                                                                         \\
$h_1(1415)$      ~~~&$1 ^1P_{1}$                                                                 ~~~&$1^{+-}$    ~~~&1416      ~~~& 1488 ~~~&$s\bar{s}$                                                                                         \\
$f_2'(1525)$     ~~~&$1 ^3P_{2}$                                                                 ~~~&$2^{++}$    ~~~&1525      ~~~& 1488 ~~~&$s\bar{s}$                                                                                         \\
\midrule[1.0pt]\midrule[1.0pt]
\end{tabular}
\end{center}
\end{table*}

\section{discussions}\label{DIS}

In this work we only focus on the states which can be
approximatively considered as a pure $s\bar{s}$ state.
Considering the fact that for the low-lying pseudoscalar
isoscalar states with $J^{PC}=0^{-+}$ there may exist a strong flavor mixing between
$n\bar{n}=(u\bar{u}+d\bar{d})/\sqrt{2}$ and $s\bar{s}$
~\cite{Godfrey:1985xj,Barnes:1996ff,Vijande:2004he,Ricken:2003ua,Klempt:2007cp,Yu:2011ta},
we omit the discussions about these states in present work.

\subsection{ Well-established vector $s\bar{s}$ states}

\subsubsection{$\phi(1020)$}

The $\phi(1020)$ resonance, as the lowest $S$-wave vector $s\bar{s}$ state $1^3S_1$, was first observed in a bubble chamber experiment at Brookhaven in 1962~\cite{Bertanza:1962zz}. In a previous study with the standard relativistic phase space the width of $\phi \to KK$ was predicted to be $\Gamma[\phi(1020) \to KK]\simeq 2.5$ MeV~\cite{Barnes:2002mu}, which is clearly smaller than the measured value
$3.5$ MeV~\cite{Tanabashi:2018oca}. To include some relativistic corrections to the phase space, we adopt the ``mock meson" method in our calculations. The partial decay width of $\phi \to KK$ is predicted to be $\Gamma[\phi(1020) \to KK]\simeq 4.1$ MeV, our result is in good
agreement with the measured value $3.5$ MeV~\cite{Tanabashi:2018oca}.

\subsubsection{$\phi(1680)$}

As the $2^3S_1$ $s\bar{s}$ state, the $\phi(1680)$ was first discovered in $e^+e^- \to K_SK^\pm\pi^\mp$~\cite{Mane:1982si}.
Both the mass and width can be well understood within the quark model. Our predictions
of the strong decay properties are shown in Table~\ref{decay1}.
It shows that the predicted decay width $\Gamma_{total}=167$ MeV is in agreement with the measured
value $150\pm 50$ MeV~\cite{Tanabashi:2018oca}. Furthermore, our calculation shows that the decay
of $\phi(1680)$ is governed by the $KK^*(892)$ mode, its branching fraction can reach up to $81\%$, which is close to the other
predictions~\cite{Barnes:2002mu,Pang:2019ttv}. In addition, our predicted partial width ratio
\begin{equation}\label{eq4}
\frac{\Gamma (KK)}{\Gamma (KK^*)}\approx0.06
\end{equation}
is comparable with the DM1 measured value $0.07\pm0.01$~\cite{Mane:1982si}. Our prediction of
\begin{equation}\label{eq4}
R_{\eta\phi/KK^*}=\frac{\Gamma (\eta\phi)}{\Gamma (KK^*)}\approx0.20
\end{equation}
is consistent with the predictions in Refs.~\cite{Barnes:2002mu,Pang:2019ttv,deQuadros:2020ntn},
however, it is about two times smaller than the measured value 0.37 from the $BaBar$ Collaboration~\cite{Aubert:2007ym}.
To clarify the inconsistency in the ratio $R_{\eta\phi/KK^*}$,
more accurate measurements are expected to be carried out in future experiments.

\subsection{$1P$-wave $s\bar{s}$ states}

%Several $1P$-wave $s\bar{s}$ states may be found in experiments.

\subsubsection{$f_2'(1525)$}

The $f_2'(1525)$ resonance listed by the PDG is widely accepted as the $1^3P_2$ $s\bar{s}$ state.
Both the mass and decay properties can be reasonably understood in the quark model~\cite{Godfrey:1985xj,Roberts:1997kq,Barnes:2002mu}.
Considering the $f_2'(1525)$ as the $1^3P_2$ $s\bar{s}$ state,
we calculate its OZI-allowed two body strong decays by using the wave function obtained
from our potential model, our results are listed in Table~\ref{decay3}.
It is found that our predicted width
\begin{equation}\label{f2w}
\Gamma_{total} \simeq 58 \ \mathrm{MeV}
\end{equation}
is sightly smaller than the average data $\Gamma_{exp.}=86\pm 5$ MeV from the PDG~\cite{Tanabashi:2018oca}.
The decay of $f_2'(1525)$ is governed by the
$KK$ mode, its branching fraction can reach up to
\begin{equation}\label{f2}
Br[f_2'(1525)\to KK]\simeq 70\%,
\end{equation}
which is close to the value 76\% predicted in Refs.~\cite{Roberts:1997kq,Barnes:2002mu} and measured value $88\%$ from the PDG~\cite{Tanabashi:2018oca}.
Furthermore, the branching fraction for the $\eta\eta$ channel is predicted to be
\begin{equation}\label{f2}
Br[f_2'(1525)\to \eta\eta]\simeq 9\%.
\end{equation}
We also find that the branching fraction ratio
\begin{equation}\label{rat}
R_{\eta\eta/KK}=\frac{\Gamma (\eta\eta)}{\Gamma (KK)}\approx 13\%
\end{equation}
is close to the average value $11.5\%$ from the PDG~\cite{Tanabashi:2018oca}.

It should be pointed out the
$KK^*(892)$ is another important decay mode of $f_2'(1525)$, the branching fraction
may reach up to
\begin{equation}\label{f2}
Br[f_2'(1525)\to KK^*(892)]\simeq 21\%.
\end{equation}
A fairly large decay rate into the $KK^*(892)$ final state is also predicted in Refs.
~\cite{Barnes:2002mu,Ye:2012gu}. This important decay mode is hoped to be measured in
future experiments. A small $n\bar{n}=(u\bar{u}+d\bar{d})/\sqrt{2}$ component may exist in the $f_2'(1525)$
resonance for a tiny decay rate into $\pi\pi$ decay channel~\cite{Ye:2012gu}.

%However, in Ref.~\cite{Geng:2008gx} the $f_2'(1525)$ was suggested to be a dynamically generated
%state from the vector-vector interaction.

\subsubsection{$h_1(1415)$}

The $h_1(1415)$ resonance is a convincing candidate for the $1^1P_1$ $s\bar{s}$ state in
the quark model~\cite{Vijande:2004he}. The recent BESIII measurements largely improved the
accuracy of the observed mass and width of $h_1(1415)$ by using the $\chi_{cJ}$~\cite{Ablikim:2015lnn} and $J/\psi$~\cite{Ablikim:2018ctf} decays.
The most precise mass and width of $h_1(1415)$ are measured to be $M=(1423.2\pm 9.4)$ MeV and
$\Gamma=(90.3\pm 27.3)$ MeV, respectively~\cite{Ablikim:2018ctf}. Our predicted mass $M=1462$ MeV together with other quark
model predictions for the $1^1P_1$ $s\bar{s}$ state (see Table~\ref{mass}) is consistent with the observation.
Due to a strong suppression by the phase space factor, $KK^*(892)$ is the only dominant decay mode.
Considering the $h_1(1415)$ as the $1^1P_1$ $s\bar{s}$ state,
with the physical mass $M=1416$ MeV the total width is predicted to be
\begin{equation}\label{h1}
\Gamma_{total}\simeq\Gamma[h_1(1415)\to KK^*(892)]\simeq 141 \ \mathrm{MeV},
\end{equation}
which is comparable with the newest data $(90.3\pm 27.3)$ MeV
from the BESIII Collaboration~\cite{Ablikim:2018ctf}, and the average data $(90\pm 15)$ MeV from the PDG~\cite{Tanabashi:2018oca}.

It should be mentioned that in some works the $h_1(1415)$ was suggested to be
a dynamically generated resonance~\cite{Jiang:2019ijx,Roca:2005nm},
or the triangle singularity might be relevant here~\cite{Guo:2017jvc}.

\subsubsection{ The $1^3P_1$ $s\bar{s}$ state}

The situation for the $1^3P_1$ $s\bar{s}$ state is ambiguous.
In theory, its mass is estimated to be $\sim 1.4-1.5$ GeV~\cite{Ishida:1986vn,Godfrey:1985xj,Vijande:2004he,Ebert:2009ub}.
In this mass range there are two candidates $f_1(1420)$ and $f_1(1510)$ for the $1^3P_1$ $s\bar{s}$ state
although the $f_1(1510)$ remains to be firmly established.
There are some long-standing puzzles about the nature of these $1^{++}$ isovector meson~\cite{Close:1997nm,Godfrey:1998pd,Guo:2017jvc,Tanabashi:2018oca}.

Considering the $f_1(1420)$ as the $1^3P_1$ $s\bar{s}$ state, our predicted mass $M=1492$ MeV
is about $70$ MeV larger than the measured value $1426$ MeV~\cite{Tanabashi:2018oca}. In Ref.~\cite{Close:1997nm},
it is mentioned that the physical resonance could be shifted to the $f_1(1420)$ mass due to the presence of the
$KK^*(892)$ threshold through a mechanism similar to that suggested in~\cite{Tornqvist:1995kr}.
On the other hand we analyze the decay properties of $f_1(1420)$ as
assignment of the $1^3P_1$ state. Due to a strong suppression by the phase space factor, $KK^*(892)$ is the only dominant decay mode.
The decay width is predicted to be
\begin{equation}\label{h1}
\Gamma_{total}\simeq \Gamma[f_1(1420)\to KK^*(892)]\simeq  296 \ \mathrm{MeV},
\end{equation}
which is too large to be comparable with the measured width $\Gamma_{exp.}\simeq 55$ MeV~\cite{Tanabashi:2018oca}.
Our conclusion is consistent with the study in Ref.~\cite{Barnes:2002mu}. It should be mentioned that
the resonance envelope may be distorted by the nearby $KK^*(892)$ threshold,
which may lead to a strong suppression of the resonance width~\cite{Barnes:2002mu}.
To know the $KK^*(892)$ threshold effects on the width, more theoretical studies are needed.
In Ref.~\cite{Chen:2015iqa}, the $f_1(1420)$ was suggested to be a mixed $1^3P_1$
state containing sizeable $n\bar{n}$ components.
Moreover, some unconventional explanations, such as a $KK^*$ molecule~\cite{Longacre:1990uc},
hybrid state~\cite{Ishida:1989xh}, or the manifestation of the $KK^*$ and $\pi a_0(980)$ decay
modes of the $f_1(1285)$~\cite{Debastiani:2016xgg} etc., are also proposed in the literature.
More discussions on $f_1(1420)$ can be found in Refs.~\cite{Tanabashi:2018oca,Close:1997nm,Godfrey:1998pd}.

The $f_1(1510)$ competes with the $f_1(1420)$ to be the $1^3P_1$ $s\bar{s}$ state~\cite{Tanabashi:2018oca}.
Considering the $f_1(1510)$ as the $1^3P_1$ $s\bar{s}$ state, our predicted mass $M=1492$ MeV
is very close to the measured average value $1518\pm 5$ MeV~\cite{Tanabashi:2018oca}.
In this case the decay width is predicted to be
\begin{equation}\label{h1}
\Gamma_{total}\simeq \Gamma[f_1(1510)\to KK^*(892)]\simeq  383 \ \mathrm{MeV},
\end{equation}
which is also too large to be comparable with the measured width $\Gamma_{exp.}\simeq 73\pm 25$ MeV
~\cite{Tanabashi:2018oca}. The $f_1(1510)$ may be not well established~\cite{Close:1997nm}.
If the $f_1(1510)$ is well established, it may not be identified as the $1^3P_1$ $s\bar{s}$ state
for its narrow decay width.

The world's largest $J/\psi$ samples at BESIII may offer an opportunity
for establishing the $1^3P_1$ $s\bar{s}$ state by observing the $J/\psi$ radiative decays
$J/\psi\to \gamma X, X\to KK^*(892)$.

\subsubsection{ The $1^3P_0$ $s\bar{s}$ state}

The $1^3P_0$ $s\bar{s}$ state is still not established. The identification of
the isoscalar $0^{++}$ mesons observed from experiments is a long-standing puzzle~\cite{Tanabashi:2018oca}.
In the quark potential model, the $1^3P_0$ state should be the lightest state
in the three $1^3P_J$ states for a strong negative tensor interaction, which has been confirmed
in the observations of the $\chi_{cJ}(1P)$ and $\chi_{bJ}(1P)$ ($J=0,1,2$) states~\cite{Tanabashi:2018oca}.
The mass splitting between $\chi_{c2}(1P)$ and $\chi_{c0}(1P)$ can reach up to $\sim 150$ MeV~\cite{Tanabashi:2018oca}.
Considering this fact, we may conclude that the mass of the $1^3P_0$ $s\bar{s}$ state should be obviously lighter than the mass of
the $f_2'(1525)$ state ($1^3P_2$ $s\bar{s}$ state), which indicates that the $f_0(1500)$ and $f_0(1710)$ resonances listed by
the PDG~\cite{Tanabashi:2018oca} cannot be identified as the $1^3P_0$ $s\bar{s}$ state.

Our predicted mass for the $1^3P_0$ $s\bar{s}$ state is $\sim 1370$ MeV,
which is consistent with the predictions in Refs.~\cite{Godfrey:1985xj,Xiao:2019qhl,Vijande:2004he,Ebert:2009ub}.
Concerning the mass, the $f_0(1370)$ is likely a candidate for the $1^3P_0$ $s\bar{s}$ state.
However, the $f_0(1370)$ decays mostly into pions ($2\pi$ and $4\pi$), this fact suggests it
more favors an $n\bar{n}$ structure~\cite{Tanabashi:2018oca}. It should be mentioned that
the experimental situation about $f_0(1370)$ is rather fluid~\cite{Tanabashi:2018oca}.
The $f_0(1370)$ resonance may actually correspond to two different states with dominant $n\bar{n}$ and $s\bar{s}$ contents, respectively~\cite{Black:1999yz,Vijande:2004he}. Thus, some resonances with a mass around $1370$ MeV observed in
the $KK$ channel might be good candidates for the $1^3P_0$ $s\bar{s}$ state.

With our predicted mass $M=1373$ MeV and wave function from the potential model,
we study the strong decay properties of the $1^3P_0$ $s\bar{s}$ state.
It is found that this state has a broad width
\begin{equation}\label{h1}
\Gamma_{total}\simeq 338 \ \mathrm{MeV},
\end{equation}
and dominantly decays into the $KK$ and $\eta\eta$ final states
with branching fractions $\sim 83\%$ and $\sim 17\%$, respectively.
The partial width ratio between $\eta\eta$ and $KK$ is predicted to be
\begin{eqnarray}\label{f0}
R_{\eta\eta/KK}=\frac{\Gamma(\eta\eta)}{\Gamma(KK)}\simeq 0.20,
\end{eqnarray}
which can be tested in future experiments.
Recently, a scalar resonance $f_0(1370)$ has been established in the $KK$ final state
from the $J/\psi\to \gamma K^+K^-,K_S^0K_S^0$ processes by using the data from
CLEO~\cite{Dobbs:2015dwa} and BESIII~\cite{Ablikim:2018izx}, respectively.
The measured mass and width of $f_0(1370)$ are  $M=1350\pm 9^{+12}_{-2}$ MeV
and $\Gamma=231\pm 21^{+28}_{-48}$ MeV, respectively~\cite{Ablikim:2018izx}.
Both our predicted mass and width for the $1^3P_0$ $s\bar{s}$ state are consistent with
the observations from CLEO and BESIII. One point should be emphasized that no obvious evidence
of $f_0(1370)$ is observed in the $\pi\pi$ spectra of the $J/\psi\to \gamma \pi^+\pi^-,\pi^0\pi^0$
processes~\cite{Dobbs:2015dwa}, which indicates that the scalar resonance $f_0(1370)$
may be dominated by the $s\bar{s}$ component.

Our predicted mass and width for
the $1^3P_0$ $s\bar{s}$ state are also consistent with the observations
extracted from the $\eta\eta$ final state~\cite{Tanabashi:2018oca}.
Thus, the scalar resonance $f_0(1370)$ observed in the $KK$ and $\eta\eta$ final states may
correspond to the $1^3P_0$ $s\bar{s}$ state. The recent analysis in Ref.~\cite{Fariborz:2015era}
also supports the $f_0(1370)$ to be an $s\bar{s}$ dominant state. A flavor mixing between $s\bar{s}$ and $n\bar{n}$
may occur in $f_0(1370)$~\cite{Ricken:2003ua,Fariborz:2015era}, which may affect our predictions.
Our conclusion can be tested by measuring the partial width ratio $R_{KK/\eta\eta}$ between $KK$ and $\eta\eta$ with a combined
study of the reactions $J/\psi\to \gamma KK,\gamma\eta\eta$ in future experiments.

\subsection{$1D$-wave states}\label{1dwave}

\subsubsection{$\phi_3(1850)$}

The $\phi_3(1850)$ resonance was first found in the $K\bar{K}$ invariant mass spectrum in the reaction
$K^-P\to K\bar{K}\Lambda$ at CERN with a mass of $M=1850\pm10$ MeV and width of $\Gamma=80^{+40}_{-30}$ MeV~\cite{AlHarran:1981tk}.
Its spin-parity was determined to be $J^P=3^{--}$ in later CERN $\Omega$~\cite{Armstrong:1982jj} and SLAC LASS~\cite{Aston:1988rf}
experiments. Besides the $K\bar{K}$ decay mode, another decay mode $K\bar K^*+c.c.$ was established in the SLAC LASS
experiment~\cite{Aston:1988rf}. The $\phi_3(1850)$ is assigned to the $1^3D_3$ $s\bar{s}$ state in the quark model~\cite{Tanabashi:2018oca}.

We study the strong decay properties of $\phi_3(1850)$ as a candidate for $\phi(1^3D_3)$.
Our results are listed in Table~\ref{decay61}. The width of $\phi_3(1850)$ is predicted to be $\Gamma\simeq 87$ MeV, which is consistent with
the experimental observations. The $\phi_3(1850)$ dominantly decays into the $K^*K^*$, $KK^*$, and $KK$ final states
with branching fractions $\sim 41\%$, $\sim 32\%$, and $\sim 23\%$, respectively.
Our predicted partial width ratio between the $KK^*$ and $KK$ channels,
\begin{equation}\label{ratiop}
R_{KK^*/KK}=\frac{\Gamma (KK^*)}{\Gamma (KK)}\approx 1.39,
\end{equation}
is close to the upper limit of the measured value $R_{KK^*/KK}^{exp.}=0.55^{+0.85}_{-0.45}$
from the LASS experiment~\cite{Aston:1988rf}, but is obviously larger than the values $\sim 0.1-0.6$ predicted in Refs.~\cite{Godfrey:1985xj,Barnes:2002mu,deQuadros:2020ntn}.

To get more knowledge of $\phi_3(1850)$, the missing dominant $K^*K^*$ decay mode
and precise branching ratios between these main decay modes are waiting to be measured in future experiments.

\subsubsection{$\phi_1(1^3D_1)$ }\label{1dwave}

The $\phi_1(1^3D_1)$ ($1^3D_1$ $s\bar{s}$) state is still not established in experiments.
Its mass is predicted to be in the range of $\sim 1.75-1.90$ GeV in various quark models~\cite{Xiao:2019qhl,Ebert:2009ub,Godfrey:1985xj,Ishida:1986vn}.
The mass of $\phi(1^3D_1)$ should be smaller than $\phi_3(1850)$ ($1^3D_3$)
because of a more negative tensor interaction contribution. With our predicted mass $M=1806$ MeV and wave function
of the $\phi_1(1^3D_1)$ state from the potential model, we study its strong decays, our results
are listed in Table~\ref{decay61}. It is found that the $\phi(1^3D_1)$ state may be a
very broad state with a width of
\begin{equation}\label{d1}
\Gamma_{total}\simeq 707 \ \mathrm{MeV}.
\end{equation}
The decays of $\phi_1(1^3D_1)$ are governed by the $KK_1(1270)$ channel.
Its partial width and  branching fraction are predicted to be
\begin{eqnarray}\label{phi1806}
\Gamma[\phi_1(1^3D_1)\to KK_1(1270)] \simeq  620 \ \mathrm{MeV},\\
Br[\phi_1(1^3D_1)\to KK_1(1270)]\simeq  88\%.
\end{eqnarray}
The other two main decay modes $KK$ and $KK^*$ have a comparable branching fraction of $\sim 4-6\%$.
Our predictions are consistent with those in Refs.~\cite{Barnes:2002mu,Pang:2019ttv}.
The $\phi_1(1^3D_1)$, as a very broad state, might be hard to observe in experiments.

Finally, it should be pointed out that the partial width of the $KK_1(1270)$ channel strongly depends on the sign
of the mixing angle of the mixed state
$|K_1(1270)\rangle =\cos\theta_{1P}|K(1^{1}P_{1})\rangle+\sin\theta_{1P}|K(1^{3}P_{1})\rangle$.
In present work we take the mixing angle as $\theta_{1P}\simeq + 45^\circ$ determined
by the decay properties of both $K_1(1270)$ and $K_1(1400)$~\cite{Tanabashi:2018oca,Blundell:1995au,Pang:2017dlw}.
However, in the quark potential model the spin-orbit interactions may result in a negative mixing angle
~\cite{Blundell:1995au,Ebert:2009ub}. In this case, the partial width of $\Gamma[KK_1(1270)]$
is a small value $\sim 10$ MeV, and the $\phi_1(1^3D_1)$ state is a fairly narrow state with
a width of $\Gamma\sim 100$ MeV. Thus, the width of $\phi_1(1^3D_1)$ predicted in theory strongly
depends on our knowledge about the $K_1(1270)$ resonance.

\subsubsection{$\phi_2(1^3D_2)$ }

The $\phi_2(1^3D_2)$ ($1^3D_2$ $s\bar{s}$) state remains to be found in experiments.
Its mass is predicted to be in the range of $\sim 1.8-1.9$ GeV in various quark models~\cite{Xiao:2019qhl,Ebert:2009ub,Godfrey:1985xj,Ishida:1986vn}.
With our predicted mass 1840 MeV for $\phi_2(1^3D_2)$, we analyze its strong decay properties (see Table~\ref{decay61}).
It is shown that the $\phi(1^3D_2)$ state has
a width of
\begin{equation}\label{d2}
\Gamma_{total}\simeq 128 \ \mathrm{MeV},
\end{equation}
which is about a factor of 2 smaller than that predicted in Refs.~\cite{Barnes:2002mu,Guo:2019wpx}.
The decays of $\phi_2(1^3D_2)$ are governed by the $KK^*(892)$ mode with a large
branching fraction
\begin{equation}\label{d2b}
Br[\phi_2(1^3D_2)\to KK^*(892)]\simeq 69\%,
\end{equation}
which is close to the predictions in Refs.~\cite{Barnes:2002mu,Guo:2019wpx}. Furthermore,
the $\phi(1^3D_2)$ may have a sizeable decay rate into the $\phi \eta$ final state.
Our predicted branching fraction is
\begin{equation}\label{d2b2}
Br[\phi_2(1^3D_2)\to \phi\eta]\simeq 21\%.
\end{equation}
The $\phi(1^3D_2)$ state may be established in the $\phi \eta$ mass spectrum.

\subsubsection{$\eta_{s2}(1^1D_2)$}

The $\eta_{s2}(1^1D_2)$ ($1^1D_2$ $s\bar{s}$) state remains to be established in experiments.
In theory, its mass is predicted to be very close to that of $\phi_3(1850)$.
The mass difference between $\eta_{s2}(1^1D_2)$ and $\phi_3(1850)$ is only several MeV. Our strong decay analysis
(see Table~\ref{decay61})
shows that the $\eta_{s2}(1^1D_2)$ mainly decays into the $KK^*$ and $K^*K^*$ channels with a
fairly narrow width $\Gamma\simeq 81$ MeV. With our predicted mass 1825 MeV for
$\eta_{s2}(1^1D_2)$, the branching fractions for the $KK^*$ and $K^*K^*$ channels are estimated to be
\begin{eqnarray}\label{phi1806}
Br[\eta_{s2}(1^1D_2)(1825)\to KK^*]\simeq 89\%,\\
Br[\eta_{s2}(1^1D_2)(1825)\to K^*K^*]\simeq 10\%.
\end{eqnarray}
Our predicted strong decay properties are consistent with those predictions in
Refs.~\cite{Barnes:2002mu}. The $1^1D_2$ quarkonium states
might be hard to produce in experiments, since no $1^1D_2$ states are established in the $s\bar{s}$,
$c\bar{c}$, and $b\bar{b}$ families.

The $\eta_{2}(1870)$ resonance with a mass of $M=1842\pm 8$ MeV and a width of $\Gamma=225\pm 14$ MeV
listed by the PDG~\cite{Tanabashi:2018oca} might be a candidate for the $\eta_{s2}(1^1D_2)$ state with a small mixing with
the $n\bar{n}$ component~\cite{Wang:2014sea}.
However, with the $\eta_{s2}(1^1D_2)$ assignment, the missing dominant
$KK^*$ decay mode in observations is hard to understand; furthermore our predicted width $\Gamma\sim 100$ MeV is
about a factor 2 smaller than the measured value $225\pm 14$ MeV. Some reviews of the $\eta_{2}(1870)$
can be found in Refs.~\cite{Barnes:2002mu,Wang:2014sea,Li:2009rka,Bing:2013fva}. To clarify the nature of
$\eta_{2}(1870)$ and establish the $\eta_{s2}(1^1D_2)$ state, the $KK^*$ and/or $K^*K^*$ final states are worth observing in
future experiments.

\subsection{$2P$-wave states}

\subsubsection{ The $2^3P_0$ $s\bar{s}$ state}

The $2^3P_0$ $s\bar{s}$ state remains to be established. In theory, its mass is predicted to be
$\sim 2.0$ GeV~\cite{Ishida:1986vn,Xiao:2019qhl,Ebert:2009ub,Godfrey:1985xj}.
With our predicted mass of 1971 MeV, we analyze
its strong decay properties (see Table~\ref{decay3}). It is found that this state might be very broad state
with a width of
\begin{equation}\label{h1}
\Gamma_{total}\simeq 849 \ \mathrm{MeV}.
\end{equation}
The $KK(1460)$, $K^*K^*$, and $KK_1(1270)$ are the dominant decay modes,
their branching fractions are predicted to be $\sim35\%$, $\sim8\%$, and $\sim51\%$, respectively.
Our predictions are consistent with those predicted in Ref.~\cite{Barnes:2002mu}.
It should be pointed out that the partial width for the $KK_1(1270)$ mode is sensitive to
the sign of the mixing angle of $K_1(1270)$. If one takes
a negative mixing angle, the partial width of $KK_1(1270)$ is very small.
Thus, our predicted strong decay properties of the $2^3P_0$ $s\bar{s}$ state
strongly depend on our knowledge of $K_1(1270)$.

Concerning the mass, the $f_{0}(2020)$ with a mass of $M=1992\pm 16$ MeV and
width of $\Gamma=442\pm 62$ MeV listed by the PDG~\cite{Tanabashi:2018oca} might be a candidate for the $2^3P_0$ $s\bar{s}$ state.
However, the $\pi\pi$, $\rho\rho$, and $\omega\omega$ decay modes seen in experiments are
not typical decay modes of the $2^3P_0$ $s\bar{s}$ state.
On the other hand, our predicted width $\Gamma_{total}=945$ MeV seems too large
to be comparable with the observation. The $f_{0}(2020)$ might be a candidate
for the $3^3P_0$~\cite{Ebert:2009ub} or $4^3P_0$~\cite{Vijande:2004he} $n\bar{n}$ state.
Furthermore, the flavor mixing between $s\bar{s}$ and $n\bar{n}$ may occur in $f_{0}(2020)$~\cite{Vijande:2004he}.
To clarify the nature of $f_{0}(2020)$ and look
for the missing $2^3P_0$ $s\bar{s}$ state, the $KK_1(1270)$, $KK(1460)$, and $K^*K^*$ final states are worth
observing in future experiments. It should be mentioned that there may exist
a great challenge to establish the missing $2^3P_0$ $s\bar{s}$ state in experiments
for its rather broad width.

\subsubsection{ The $2^3P_2$ $s\bar{s}$ state}

The $2^3P_2$ $s\bar{s}$ state remains to be established. In theory, its mass is predicted to be
$\sim 2.0$ GeV~\cite{Ishida:1986vn,Xiao:2019qhl,Ebert:2009ub,Godfrey:1985xj}.
Our quark model predicted value is $M=2030$ MeV. According to our analysis of the strong decay properties (see Table~\ref{decay3}),
this state might have a moderate width of
\begin{equation}\label{3p2}
\Gamma_{total}\simeq 147 \ \mathrm{MeV},
\end{equation}
and dominantly decays into the $K^*K^*$, $KK_2^*(1430)$, $KK_1(1270)$, and $KK_1(1400)$ final states
with branching fractions $\sim17\%$, $\sim24\%$, $\sim18\%$, and $\sim29\%$, respectively.
The $2^3P_2\to KK$ is a $D$-wave suppression process, the branching ratio is only $\sim 3\%$.
%Our predictions are obviously different from those in Ref.~\cite{Barnes:2002mu}.

Some signals of the $2^3P_2$ $s\bar{s}$ state might have been observed in experiments.
The $f_2(2010)$ with a mass $M=2011^{+62}_{-76}$ MeV and
width $\Gamma=202^{+67}_{-62}$ MeV listed by the PDG~\cite{Tanabashi:2018oca} might be a good candidate for the $2^3P_2$
$s\bar{s}$ state. The $f_2(2010)$ resonance was extracted by a partial wave analysis of the $KK$ and
$\phi\phi$ mass spectra of the reactions $\pi^-p\to \phi\phi n, KK n$~\cite{Tanabashi:2018oca}.
Recently, the $f_2(2010)$ resonance was also observed in $J/\psi\to \gamma \phi\phi$ at BESIII~\cite{Ablikim:2016hlu}.
As an assignment of the $2^3P_2$ $s\bar{s}$ state, our predicted mass and width of $f_2(2010)$ are in good agreement
with the observations. It should be mentioned that although the mass of $f_2(2010)$ lies under the $\phi\phi$
threshold, as the $2^3P_2$ $s\bar{s}$ state, it can contribute to the $\phi\phi$ mass spectrum
due to a sizeable coupling to $\phi\phi$~\cite{Barnes:2002mu}. Thus, the $KK$ and
$\phi\phi$ decay modes of $f_2(2010)$ observed in experiments are also consistent with the predictions.

However, in Ref.~\cite{Ye:2012gu} the $f_2(1810)$ listed by the PDG~\cite{Tanabashi:2018oca} was suggested to be a
$2^3P_2$ $s\bar{s}$ dominant state with a few $n\bar{n}$ components. Obviously, the observed mass of
$f_2(1810)$ is too small to be comparable with predictions from most of the quark models.
To definitively establish the $2^3P_2$ $s\bar{s}$ state, the dominant decay modes
$KK_1(1270)$, $K^*K^*$, and $KK_2^*(1430)$ are worth observing in future experiments.

\subsubsection{ The $2^3P_1$ $s\bar{s}$ state}

There is no hint about the $2^3P_1$ $s\bar{s}$ state from experiments.
Our predicted mass for this state is $M=2030$ MeV, which is in good agreement with the other predictions~\cite{Ishida:1986vn,Xiao:2019qhl,Ebert:2009ub,Godfrey:1985xj}.
According to our analysis of the strong decay properties (see Table~\ref{decay3}),
this state might have a broad width of
\begin{equation}\label{3p1}
\Gamma_{total}\simeq 315 \ \mathrm{MeV}.
\end{equation}
The $2^3P_1$ $s\bar{s}$ state dominantly decays into the $KK_2^*(1430)$, $KK_1(1270)$, $K^*K^*$, and $KK^*$,
channels with branching fractions $\sim35\%$, $\sim29\%$, $\sim16\%$, and $\sim13\%$ respectively.
The dominant decay modes $KK_1(1270)$, $K^*K^*$ and $KK^*$ and a broad width $\sim 300$ MeV for the $2^3P_1$ $s\bar{s}$
state were also predicted in Ref.~\cite{Barnes:2002mu}. The $2^3P_1$ $s\bar{s}$ state might have large
potentials to be established at BESIII by using the $J/\psi$ or $\psi(2S)$ decays, such as
$J/\psi\to \gamma X, X\to KK_2^*(1430), K^*K^*,KK^*$.

\subsubsection{ The $2^1P_1$ $s\bar{s}$ state}

The $2^1P_1$ $s\bar{s}$ state is still not established.
Our quark model predicted mass is $M=1991$ MeV, which is in good agreement with most quark model predictions~\cite{Ishida:1986vn,Xiao:2019qhl,Ebert:2009ub,Godfrey:1985xj}.
According to our analysis of the strong decay properties (see Table~\ref{decay3}), this state might have a moderate width of
\begin{equation}\label{2h1}
\Gamma_{total}\simeq 179 \ \mathrm{MeV},
\end{equation}
and dominantly decay into the $KK_2^*(1430)$, $K^*K^*$, and $KK^*$ final states with
branching fractions $\sim46\%$, $\sim17\%$, and $\sim18\%$, respectively.
The dominant decay modes $K^*K^*$ and $KK^*$ and a moderate width $\sim 190$ MeV for the $2^1P_1$ $s\bar{s}$
state were also predicted in Ref.~\cite{Barnes:2002mu}.

Recently, a new $1^{+-}$ resonance $X(2062)$ with a $3.8\sigma$
significance might have been observed in the $\eta' \phi$
mass spectrum of the decay $J/\psi\to \phi \eta \eta'$ at BESIII~\cite{Ablikim:2018xuz}.
The mass and width were determined to be $M=(2062.8\pm13.1\pm7.2)$ MeV and
$\Gamma=(177\pm36\pm35)$ MeV, respectively. It is interesting to find that
the observed mass and width of $X(2062)$ are very similar to our predictions
for the $2^1P_1$ $s\bar{s}$ state. In Ref.~\cite{Wang:2019qyy} the $X(2062)$ was also
suggested to be the $2^1P_1$ $s\bar{s}$ state with a small $n\bar{n}$ component.
If the $X(2062)$ correspond to the  $2^1P_1$ $s\bar{s}$ state indeed,
the branching fractions into the $\eta \phi$ and $\eta' \phi$ final states are predicted to be
\begin{eqnarray}\label{h1200}
Br[X(2062)\to \eta \phi]\simeq 4\%,\\
Br[X(2062)\to \eta' \phi]\simeq 1\%,
\end{eqnarray}
which are about one order of magnitude smaller than that into the $KK^*$ final state.
To confirm the existence of $X(2062)$, the decay processes $J/\psi\to \eta X, X\to K^*K/ \eta \phi$
are worth observing in future experiments.

\subsection{$1F$-wave states}

\subsubsection{ The $1^3F_2$ $s\bar{s}$ state}

The $1^3F_2$ $s\bar{s}$ state is not established. Our quark model predicted mass is $M=2143$ MeV,
which is in good agreement with the prediction with the covariant oscillator quark model~\cite{Ishida:1986vn},
while our result is about 100 MeV smaller than the predictions with the relativized quark model
~\cite{Xiao:2019qhl,Godfrey:1985xj} and the relativistic quark model~\cite{Ebert:2009ub}.
According to our analysis of the strong decay properties (see Table~\ref{decay1f2}),
this state might be a broad state with a width of
\begin{equation}\label{h1}
\Gamma_{total}\simeq 308 \ \mathrm{MeV}.
\end{equation}
The $1^3F_2$ $s\bar{s}$ state has relatively large decay rates into the
$K^*K^*$, $KK_2^*(1430)$, and $KK_1(1270)$ channels,
their branching fractions are predicted to be $\sim6\%$, $\sim9\%$, and $\sim62\%$, respectively.
The decay rates into $KK$, $KK^*$ are also sizeable, their branching fractions are
predicted to be $\sim1.8\%$, $\sim4\%$, respectively.
The $\eta\eta$, $\eta'\eta'$, and $\phi\phi$ channels have a comparable branching fraction
$\sim0.3\%$, and the branching fraction of $\eta\eta'$ is about $1.2\%$, they may be ideal channels
for looking for the $1^3F_2$ $s\bar{s}$ state. Our predicted strong properties are roughly consistent with those predicted in Refs.~\cite{Barnes:2002mu,Blundell:1995ev}.

Concerning the mass, the $f_2(2150)$ resonance listed by the PDG~\cite{Tanabashi:2018oca} might be a good
candidate for the $1^3F_2$ $s\bar{s}$ state. This resonance might have been observed
by some experimental groups in different reactions with a similar mass, however,
most of the extracted resonance widths are notably different.
It should be mentioned that the WA102 Collaboration established the $f_2(2150)$ resonance in both $K^+K^-$ and
$\eta\eta$ final states with consistent mass $M\simeq 2130$ MeV and width $\Gamma\simeq 270$ MeV~\cite{Barberis:1999am,Barberis:2000cd}.
Considering $f_2(2150)$ as the $1^3F_2$ $s\bar{s}$ state, our predicted mass and width are
in good agreement with the observations from the WA102 Collaboration~\cite{Barberis:1999am,Barberis:2000cd}.
Recently, the BEIII Collaboration observed a broad isoscalar $2^{++}$ state around 2.2 GeV via the
process $J/\psi\to \gamma K_SK_S$~\cite{Ablikim:2018izx}. The mass and width are determined to be $M=2233\pm 34$$^{+9}_{-25}$ MeV
and $\Gamma=507\pm 37$$^{+18}_{-21}$ MeV, respectively. Considering the uncertainties of the observations,
the resonance observed at BESIII might be an assignment of the $1^3F_2$ $s\bar{s}$ state as well.
Furthermore, some weak evidence of $f_2(2150)$ was also observed in $J/\psi\to \gamma \phi\phi$
by the BEIII Collaboration~\cite{Ablikim:2016hlu}.

Finally, it should be pointed out that in some works the $f_2(2150)$
state was assigned to the $2^3P_2$ $s\bar{s}$ state~\cite{Anisovich:2011in},
or the $3^3P_2$ $s\bar{s}$ state~\cite{Ye:2012gu}. However, with the $2^3P_2$ (or $3^3P_2$) assignment the mass
of $f_2(2150)$ is about 100 MeV larger (or 300 MeV smaller) than the theoretical predictions (see Table~\ref{mass}).
A combined analysis of the $2^{++}$ resonances around $2.1-2.4$ GeV in via $J/\psi$ radiative
decays into the $KK$, $\eta\eta$, $\eta\eta'$, and $\phi\phi$ final states might be helpful to understand
the nature of the $f_2(2150)$ resonance.

\begin{figure}[!htbp]
\begin{center}
\centering  \epsfxsize=8.2cm \epsfbox{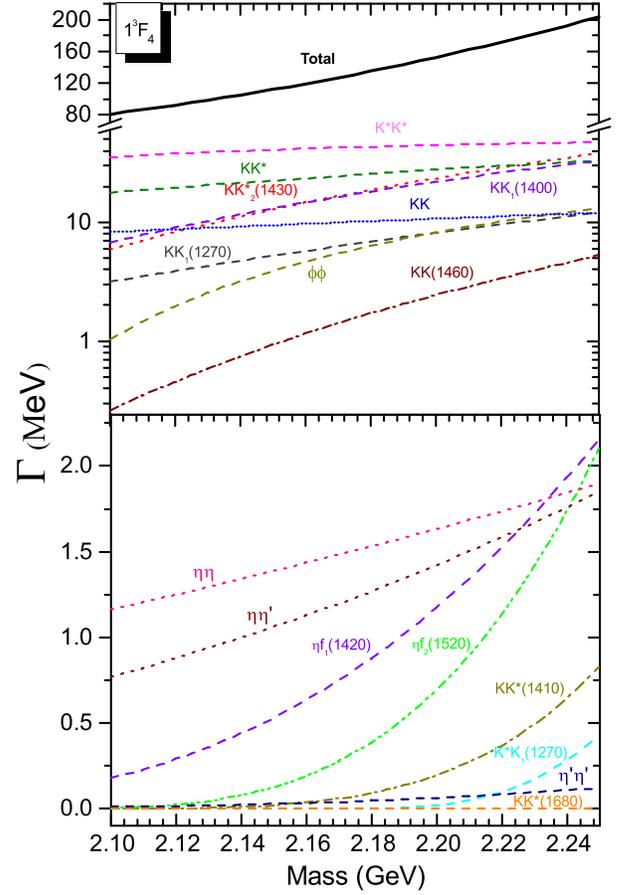}
\vspace{-0.2 cm}\caption{Variation of the decay widths for the $1^3F_4$ $s\bar{s}$ state with its mass.} \label{widthf4}
\end{center}
\end{figure}

\subsubsection{ The $1^3F_4$ $s\bar{s}$ state}

The $1^3F_4$ $s\bar{s}$ state is not established. Our predicted mass is $M=2078$ MeV, which is close to the prediction with the covariant oscillator quark model~\cite{Ishida:1986vn}, while it is about 120 MeV smaller than the predictions with the relativized quark model
~\cite{Xiao:2019qhl,Godfrey:1985xj}. The mass splitting $\Delta M$ between $1^3F_4$ and $1^3F_2$ predicted within
various quark models is very different due to the poor determination of the
spin-orbit interaction for the high spin states. With the nonrelativistic quark model
we predict $\Delta M=-68$ MeV, however, very different values $\Delta M=-40$, $+143$, $+40$ MeV
are obtained from the relativized quark model~\cite{Xiao:2019qhl,Godfrey:1985xj},
relativistic quark model~\cite{Ebert:2009ub}, and covariant oscillator quark model~\cite{Ishida:1986vn}, respectively.
According to our analysis of the strong decay properties (see Table~\ref{decay1f2}),
this state might be a relatively narrow state with a width of
\begin{equation}\label{h1}
\Gamma_{total}\simeq 70 \ \mathrm{MeV}.
\end{equation}
The main decay channels are $KK$, $KK^*$, $K^*K^*$, $KK_2^*(1430)$, $KK_1(1270)$, and $KK_1(1400)$.
Our predicted strong decay properties are consistent with those predicted in Refs.~\cite{Barnes:2002mu,Blundell:1995ev}.
Considering the fairly large uncertainties of the predicted mass of the $1^3F_4$ $s\bar{s}$ state,
we plot the decay properties as functions of the mass in Fig.~\ref{widthf4}. Some sensitivities of the
decay properties to the mass can be clearly seen from the figure. If the $1^3F_4$ $s\bar{s}$ state
has a high mass of $2.2$ GeV as that predicted in the relativized quark model~\cite{Xiao:2019qhl,Godfrey:1985xj},
the $\phi\phi$ decay mode may be an important decay mode as well.

There might have been some experimental evidence for this state.
The LASS Collaboration observed a rather narrow $4^{++}$ resonance (denoted by $f_4(2210)$)
with a mass and width of $M=2209^{+17}_{-15}$ MeV and $\Gamma=60^{+107}_{-57}$ MeV
by an analysis of the $K^+K^-$ mass spectrum from the reaction $K^-p\to K^+K^- \Lambda$~\cite{Aston:1988yp}.
The mass and width values of $f_4(2210)$ are consistent with those obtained by MARK III
for the $X(2200)$ from the reaction $J/\psi\to \gamma KK$~\cite{Baltrusaitis:1985pu}.
A similar resonance with a mass and width of $M=2231\pm 10$ MeV and $\Gamma=130\pm 50$ MeV
was also observed in the $\phi\phi$ final state by WA67 (CERN SPS)~\cite{Booth:1985kv}.
The $f_4(2210)$ might be an assignment of the $1^3F_4$ $s\bar{s}$ state~\cite{Barnes:2002mu,Blundell:1995ev}.

Considering the $f_4(2210)$ resonance as the $1^3F_4$ $s\bar{s}$ state,
we find that our predicted decay width
\begin{equation}\label{f4}
\Gamma_{total}\simeq 161 \ \mathrm{MeV}
\end{equation}
is consistent with the observed width from experiments.
The branching fractions for the dominant decay modes $KK$, $KK^*$, $K^*K^*$, $KK_2^*(1430)$, $KK_1(1270)$, and $KK_1(1400)$
are predicted to be $\sim7\%$, $\sim18\%$, $\sim28\%$, $\sim16\%$, $\sim6\%$, and $\sim15\%$, respectively.
Furthermore, it is found that the $\phi\phi$ mode also plays
an important role in the decay, its branching fraction may reach up to $\sim 6\%$,
which is comparable with that of the $KK$ mode. The partial width ratio between
$\phi\phi$ and $KK$ is predicted to be
\begin{eqnarray}\label{f4200}
R_{\phi\phi/KK}=\frac{\Gamma(\phi\phi)}{\Gamma(KK)}\simeq 0.82,
\end{eqnarray}
which can be used to test the nature of the $f_4(2210)$.
The fairly large decay rates of $f_4(2210)$ into the $KK$ and $\phi\phi$ final states
may explain why it has been seen in these channels.
To definitely establish the $1^3F_4$ $s\bar{s}$ state and confirm the nature of $f_4(2210)$
more accurate observations in the dominant decay modes are needed in future experiments.

\subsubsection{ The $1^3F_3$ $s\bar{s}$ state}

There is no hint about the $1^3F_3$ $s\bar{s}$ state from experiments.
Our quark model predicted mass is $M=2128$ MeV,
which is in good agreement with the prediction with the covariant oscillator quark model~\cite{Ishida:1986vn},
while it is about 100 MeV smaller than the predictions with the relativized quark model
~\cite{Xiao:2019qhl,Godfrey:1985xj} and the relativistic quark model~\cite{Ebert:2009ub}. The mass of $1^3F_3$ might highly overlap with that of
$1^3F_2$ with a mass splitting $\sim 10-30$ MeV~\cite{Ishida:1986vn,Xiao:2019qhl,Godfrey:1985xj}.
If $f_2(2150)$ corresponds to the $1^3F_2$ $s\bar{s}$ state indeed, the mass of the
$1^3F_3$ $s\bar{s}$ state is most likely to be in the range of $\sim2120-2160$ MeV.
According to our analysis of the strong decay properties (see Table~\ref{decay1f2}),
this state might be a broad state with a width of
\begin{equation}\label{h1}
\Gamma_{total}\simeq 245 \ \mathrm{MeV}.
\end{equation}
The dominant decay channels might be $KK_2^*(1430)$, $KK^*$, $K^*K^*$, and $KK_1(1270)$, and their branching fractions are
predicted to be $\sim56\%$, $\sim13\%$, $\sim8\%$, and $\sim15\%$, respectively.
It is interesting to find that the decay rate into the $\eta f_2'(1525)$ channel is sizeable,
the branching fraction could reach up to $\sim 6.5\%$. The $\eta f_2'(1525)$ may be a useful
channel for searching for the missing $1^3F_3$ $s\bar{s}$ state in experiments.
Our main predictions of the strong decay properties are consistent with those predicted in Ref.~\cite{Barnes:2002mu}.

\subsubsection{ The $1^1F_3$ $s\bar{s}$ state}

The $1^1F_3$ $s\bar{s}$ state is not established. Our quark model predicted mass is $M=2111$ MeV,
which is in good agreement with the prediction with the covariant oscillator quark model~\cite{Ishida:1986vn},
while it is about 100 MeV smaller than the predictions with the relativized quark model
~\cite{Xiao:2019qhl,Godfrey:1985xj} and the relativistic quark model~\cite{Ebert:2009ub}.
According to our analysis of the strong decay properties (see Table~\ref{decay1f2}),
this state might be a moderately broad state with a width of
\begin{equation}\label{h1}
\Gamma_{total}\simeq 178\ \mathrm{MeV}.
\end{equation}
The decays are governed by the $KK_2^*(1430)$ channel with a branching fraction $\sim 60\%$.
The $KK^*$ and $K^*K^*$ decay modes are another two important decay modes, they have
a comparable branching fraction of $\sim 12-18\%$. Our main predictions are consistent with those
predicted in Ref.~\cite{Barnes:2002mu}. As an attractive decay mode for observations,
the $\eta \phi$ channel may have some a sizeable branching fraction $\sim 4\%$, which is about
three times smaller than the value predicted in Ref.~\cite{Barnes:2002mu}.
%This state can be diffractively photoproduced

\subsection{$2D$-wave states}

\subsubsection{ The $2^3D_1$ $s\bar{s}$ state}

The $2^3D_1$ $s\bar{s}$ state is not established. In various quark models,
its mass is predicted to be in the range of $\sim 2.26-2.35$ GeV~\cite{Xiao:2019qhl,Ebert:2009ub,Godfrey:1985xj,Ishida:1986vn}.
Our predicted mass $M=2272$ MeV is comparable with the other model predictions within an uncertainty of about $\pm60$ MeV.
According to our analysis of the strong decay properties (see Table \ref{decay2d1}),
this state might be a broad state with a width of
\begin{equation}\label{2d1}
\Gamma_{total}\simeq 322\ \mathrm{MeV},
\end{equation}
and mainly decays into the $KK(1460)$, $K^*K^*$, $KK_2^*(1430)$, $KK_1(1270)$, $K^*K_1(1270)$, and $KK^*(1410)$
final states with branching fractions $\sim 12\%$, $\sim 9\%$, $\sim 8\%$, $\sim 27\%$, $\sim 21\%$, and $\sim 10\%$, respectively.
Furthermore, the $KK$ and $KK^*$ modes have some sizeable contributions to the decays,
they have a comparable branching fraction of $\sim2\%$. The decay properties for the $2^3D_1$ $s\bar{s}$ state
roughly agree with the predictions in Refs.~\cite{Pang:2019ttv,Ding:2007pc}.

\subsubsection{ The $2^3D_2$ $s\bar{s}$ state}

There is no hint about the $2^3D_2$ $s\bar{s}$ state from experiments.
In our nonrelativistic quark model calculation, its mass is predicted to be $M=2297$ MeV, which is slightly smaller than the previous
quark model predictions $\sim 2320-2350$ MeV~\cite{Xiao:2019qhl,Ebert:2009ub,Godfrey:1985xj,Ishida:1986vn}.
According to our analysis of the strong decay properties (see Table \ref{decay11}),
this state might be a moderately broad state with a width of
\begin{equation}\label{2d1}
\Gamma_{total}\simeq 232\ \mathrm{MeV},
\end{equation}
and mainly decays into the $KK^*(1410)$, $KK_2^*(1430)$, $K^*K_1(1270)$, and $KK_3^*(1780)$ final states.
Their branching fractions are predicted to be $\sim 30\%$, $\sim 21\%$, $\sim 10\%$, and $\sim 13\%$,
respectively. Furthermore, the $KK^*$, $K^*K^*$, and $KK_1(1400)$ modes have some sizeable contributions to the
decays with a comparable branching fraction of $\sim3-7\%$.

\subsubsection{ The $2^3D_3$ $s\bar{s}$ state}

There is no hint about the $2^3D_3$ $s\bar{s}$ state from experiments.
In our nonrelativistic quark model calculation, its mass is predicted to be $M=2285$ MeV, which is slightly
($50-80$ MeV) smaller than the previous
quark model predictions $\sim 2340-2360$ MeV~\cite{Xiao:2019qhl,Ebert:2009ub,Godfrey:1985xj,Ishida:1986vn}.
According to our analysis of the strong decay properties (see Table \ref{decay2d1}),
this state might be the narrowest state in the $2D$ states with a width of
\begin{equation}\label{2d1}
\Gamma_{total}\simeq 136\ \mathrm{MeV}.
\end{equation}
The $2^3D_3$ $s\bar{s}$ state mainly decays into the $KK(1460)$, $KK^*(1410)$, $K^*K^*$, $KK^*_2(1430)$,and $KK_1(1270)$ final
states with branching fractions $\sim 19\%$, $\sim 13\%$, $\sim 12\%$, $\sim 11\%$, and $\sim 17\%$,
respectively. Furthermore, the $KK$, $KK^*$, $KK_1(1400)$, and $K^*K_1(1270)$ modes may
have some sizeable contributions to the decays with a comparable branching fraction of $\sim3-8\%$.

\subsubsection{ The $2^1D_2$ $s\bar{s}$ state}

There is no hint about the $2^1D_2$ $s\bar{s}$ state from experiments.
In our nonrelativistic quark model calculation, its mass is predicted to be $M=2282$ MeV, which is slightly
($40-60$ MeV) smaller than the previous
quark model predictions $\sim 2320-2340$ MeV~\cite{Xiao:2019qhl,Ebert:2009ub,Ishida:1986vn}.
According to our analysis of the strong decay properties (see Table \ref{decay11}),
this state might have a moderate width of
\begin{equation}\label{2d1}
\Gamma_{total}\simeq 208 \ \mathrm{MeV},
\end{equation}
and mainly decays into the $KK^*(1410)$, $KK^*_2(1430)$, and $K^*K_1(1270)$ final states,
with a comparable branching fraction $\sim 14-27\%$. Furthermore,
the $2^1D_2$ $s\bar{s}$ state may have sizeable decay rates into the $KK^*$, $K^*K^*$,
and $KK^*_0(1430)$ final states with comparable branching fractions of $\sim5\%$.

\subsection{Possibility of the $\phi(2170)$ as a $1^{--}$ $s\bar{s}$ state}

The vector meson resonance $\phi(2170)$ (often denoted as $Y (2175)$ in the literature)
was first observed with a mass $M=2175\pm 35$ MeV and a width $\Gamma=58\pm 36$ MeV
by the $BaBar$ Collaboration in the initial state radiation (ISR) process
$e^+e^-\to \gamma_{ISR}\phi f_0(980)$~\cite{Aubert:2006bu}. In addition $BaBar$ also
observed evidence of $\phi(2170)$ in the process $e^+e^-\to \gamma_{ISR}\phi \eta$~\cite{Aubert:2007ym}.
Subsequently, the $\phi(2170)$ was confirmed in the BES experiments
$J/\psi\to \eta \phi f_0(980)$, $J/\psi\to \eta \phi \pi^+\pi^-$, and $e^+e^-\to \eta Y(2175)$~\cite{Ablikim:2007ab,Ablikim:2017auj,Ablikim:2014pfc},
and Belle experiment $e^+e^-\to \phi \pi^+\pi^-$~\cite{Shen:2009zze}.
Recently, a partial wave analysis (PWA) of the process $e^+e^-\to K^+K^-\pi^0\pi^0$ was performed
by the BESIII Collaboration, it is observed that the $\phi(2170)$ has a sizable
partial width to $K^+(1460)K^-$, $K^+_1(1400)K^-$, and $K^+_1(1270)K^-$, but a much smaller partial width to $K^{*+}(892)K^{*-}(892)$
and $K^{*+}(1410)K^{-}$~\cite{Ablikim:2020pgw}. Very recently, the $\phi(2170)$
was also clearly seen in the Born cross sections of $e^+e^-\to \phi\eta'$ by the BESIII Collaboration~\cite{1788734}.
It should be mentioned that some measurements of the processes $e^+e^-\to K^+K^-K^+K^-$~\cite{Lees:2011zi,Ablikim:2019tpp,Aubert:2005eg}, $\phi K^+K^-$~\cite{Ablikim:2019tpp}, and $K^+K^-$~\cite{Ablikim:2018iyx,BABAR:2019oes} have been carried out at $BaBar$ and BESIII, however, no significant signals of $\phi(2170)$ were found in the these reactions.

There are long-standing puzzles about the nature of $\phi(2170)$. Many interpretations, such as a
conventional $3^3S_1$ or $2^3D_1$ $s\bar{s}$ state~\cite{Barnes:2002mu,Pang:2019ttv,Ding:2007pc,meson4,Coito:2009na}, an $\bar{s}sg$ hybrid state~\cite{hybrid1,Ho:2019org},
a tetraquark state~\cite{tetra1,tetra2,tetra3,tetra4,tetra5,tetra6,tetra7,tetra8,Drenska:2008gr}, a $\Lambda\bar{\Lambda}$ bound state~\cite{Lambda1,Lambda2,Dong:2017rmg,Lambda4}, or a resonant state of the $\phi KK$ system~\cite{MartinezTorres:2008gy,GomezAvila:2007ru},
etc., have been widely discussed in the literature. However, no interpretation has yet
been established. In the following, we discuss the possibilities of $\phi(2170)$
as the conventional $3^3S_1$ and $2^3D_1$ $s\bar{s}$ states, or a mixing state between them.

\begin{figure}[!htbp]
\begin{center}
\centering  \epsfxsize=8.2cm \epsfbox{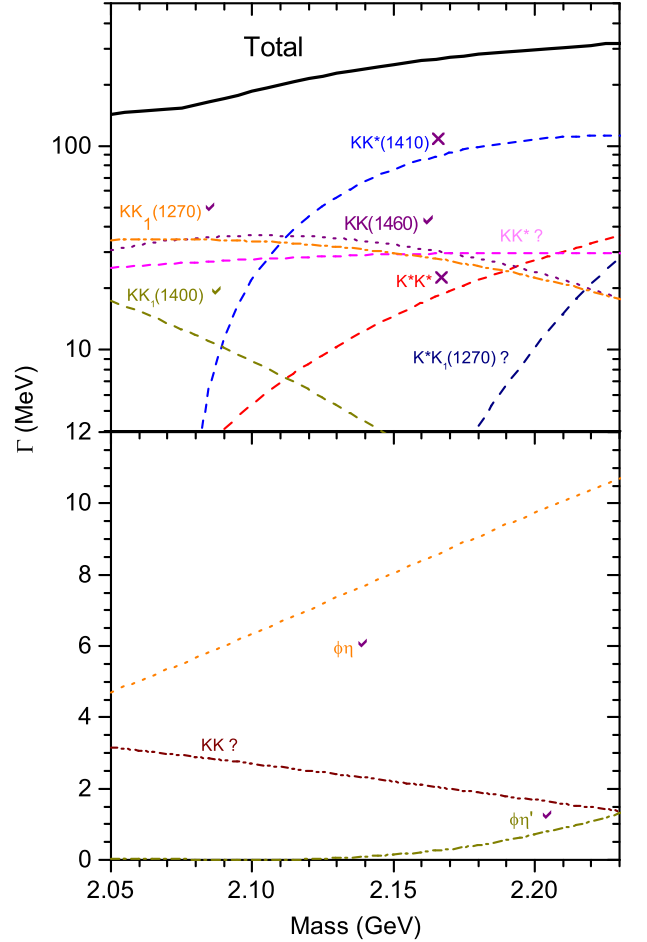}
\vspace{-0.2 cm}\caption{Variation of the decay widths for the $3^3S_1$ $s\bar{s}$ state with its mass.
The observed decay modes of $\phi(2170)$ are labeled with $``\checkmark"$, the decay modes which are not seen in experiments are
labeled with $``\times"$, and the decay modes which should be further confirmed are labeled with $``?"$.} \label{widthsw}
\end{center}
\end{figure}

\begin{figure}[!htbp]
\begin{center}
\centering  \epsfxsize=8.2cm \epsfbox{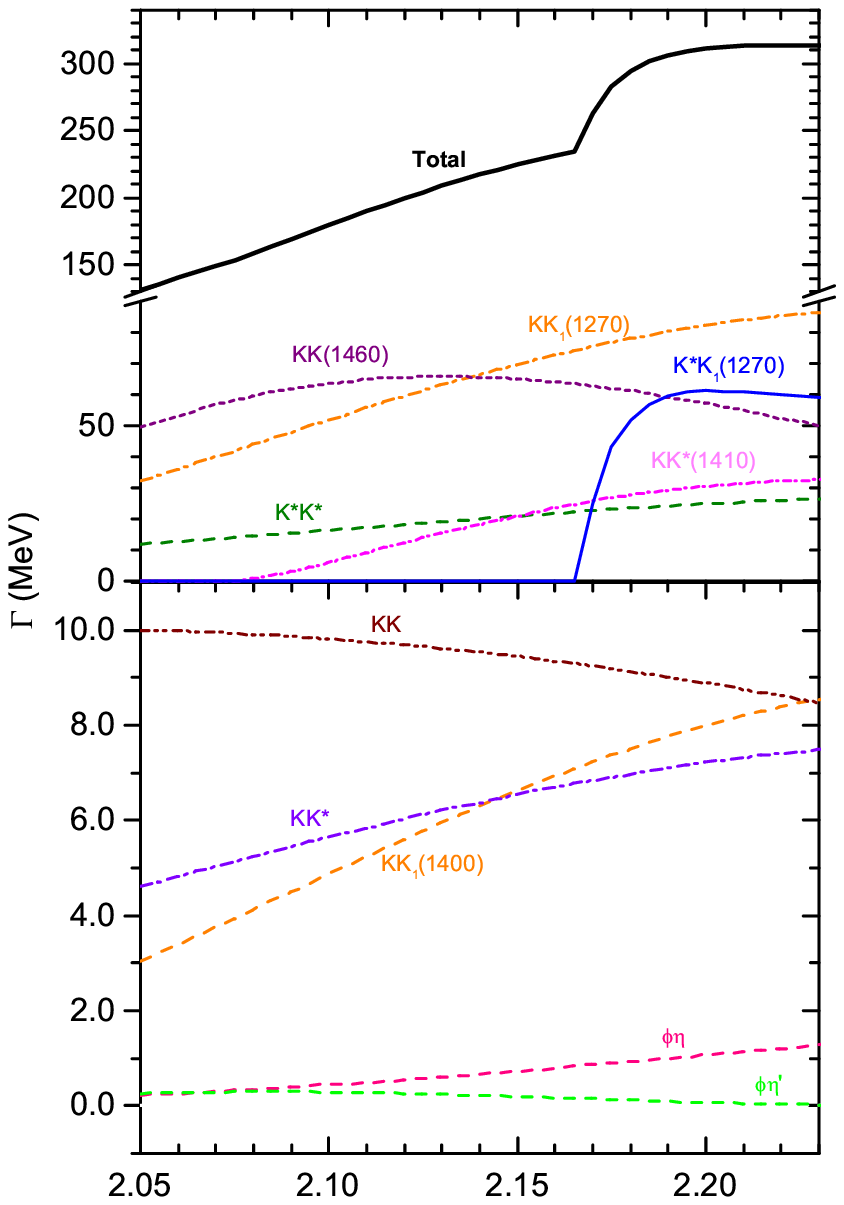}
\vspace{-0.2cm}\caption{Variation of the decay widths for the $2^3D_1$ $s\bar{s}$ state with its mass.} \label{widthdw}
\end{center}
\end{figure}

\begin{figure}[!htbp]
\begin{center}
\centering  \epsfxsize=8.2cm \epsfbox{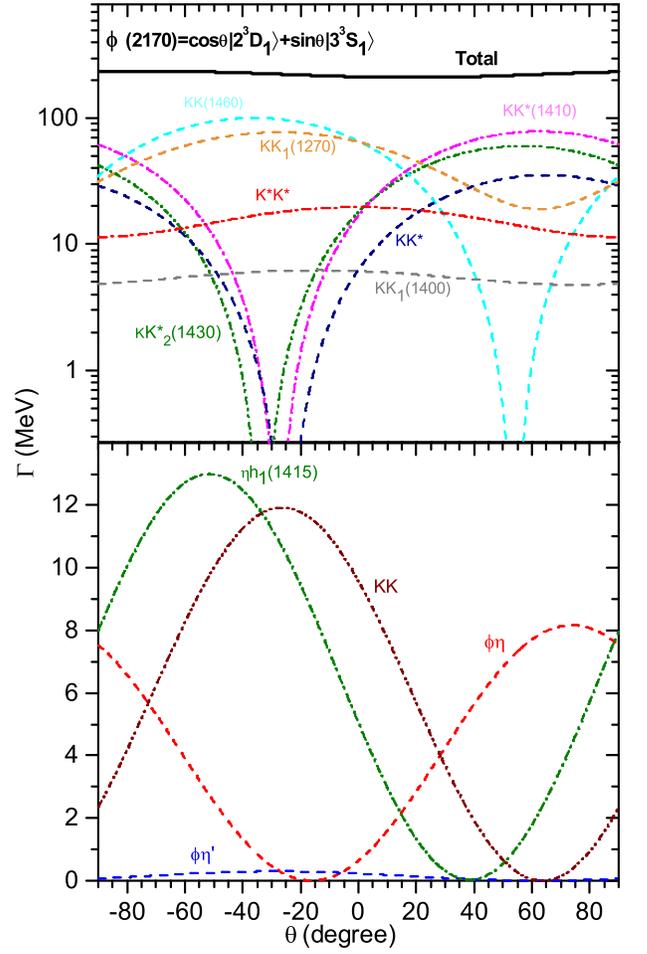}
\vspace{-0.2cm}\caption{Strong decay of $\phi(2170)$ versus the mixing angle $\theta$. } \label{widthsd}
\end{center}
\end{figure}

\subsubsection{The $\phi(3S)$ assignment}

The mass of the $\phi(3S)$ ($3^3S_1$ $s\bar{s}$) state is estimated to be in the range of $\sim2.05-2.25$ GeV
in various quark models~\cite{Ishida:1986vn,Xiao:2019qhl,Ebert:2009ub,Godfrey:1985xj,Pang:2019ttv}.
Concerning the mass, the $\phi(2170)$ is a good candidate for the $\phi(3S)$ state.
Our predicted mass $M=2198$ MeV is close to the upper limit of the observations.
It should be mentioned that the measured mass of $\phi(2170)$
has a fairly large uncertainty, which scatters in the range of $\sim 2.05-2.20$ GeV.
In this possible range we plot the strong decay properties as functions
of the initial state mass in Fig.~\ref{widthsw}. It is found that the partial widths for the $KK^*(1410)$, $K^*K^*$, $KK_1(1400)$, $KK_1(1270)$
decay modes is very sensitive to the mass of $\phi(2170)$. Specially, taking three typical
masses $2079$, $2135$, and $2175$ MeV for $\phi(2170)$,
we give our results in Table~\ref{decay2170} as well.

If the mass of $\phi(2170)$ is around $2135-2175$ MeV, as an assignment of the $\phi(3S)$ state,
the $\phi(2170)$ should be a moderately broad state with a width of $\sim 240-270$ MeV.
Although the predicted width is close to the upper limit of the observed value from Belle~\cite{Shen:2009zze},
the decay modes are inconsistent with the observations, for example the $KK^*(1410)$ and $K^*K^*$ decay modes,
as the main modes of $\phi(3S)$, were not observed in the recent BESIII experiment~\cite{Ablikim:2020pgw}.

On the other hand, if taking a smaller mass of $\sim2.08-2.1$ GeV for $\phi(2170)$,
one can find that the $\phi(2170)$ has a relatively narrow width of $\sim170-200$ MeV,
and dominantly decays into the $KK^*$, $KK(1460)$,
$KK_2^*(1430)$, $KK_1(1270)$, and $KK_1(1400)$ final states. In this case, as an assignment of $\phi(3S)$,
both the predicted width and decay modes of $\phi(2170)$ are consistent with the present observations.
The branching fractions for the interesting decay modes $\phi\eta $ and $\phi\eta'$ in experiments
are predicted to be $O(10^{-2})$ and $O(10^{-3})$, respectively.

As a whole, the possibility of $\phi(2170)$ as a candidate
for $\phi(3S)$ cannot be excluded. The decay properties of $\phi(3S)$
strongly depend on its mass. Precise measurements of branching fraction ratios between the main decay modes
and the resonance parameters are crucial to confirm whether
the $\phi(2170)$ can be assigned to the $\phi(3S)$ state or not.

\subsubsection{The $\phi(2D)$ assignment}

Concerning the mass, there is the possibility to assign $\phi(2170)$ as the $\phi(2D)$ ($2^3D_1$ $s\bar{s}$) state.
The mass of $\phi(2D)$ predicted from various quark models is $\sim 2.25-2.35$ GeV~\cite{Ishida:1986vn,Xiao:2019qhl,
Ebert:2009ub,Godfrey:1985xj,Pang:2019ttv}, which is about $100-150$ MeV higher than that of $\phi(2170)$.
Considering $\phi(2170)$ as the $2^3D_1$  $s\bar{s}$ state, we study the strong
decay properties. By varying the mass of $\phi(2170)$ in its possible range
of $\sim 2.05-2.20$ GeV, we plot the strong decay widths in Fig.~\ref{widthdw}.
Specially, taking three typical values $2079$, $2135$, and $2175$ MeV for the mass of $\phi(2170)$,
we give our predictions of the decay properties in Table~\ref{decay2170} as well.
It is found that the partial widths of $KK^*(1410)$, $KK_1(1270)$, $KK_1(1400)$,
and $K^*K^*(1270)$ are sensitive to the mass of $\phi(2170)$.

If the mass of $\phi(2170)$ is around $2175$ MeV, as an assignment of the $\phi(2D)$ state,
$\phi(2170)$ should be a broad state with a width of $\sim 300$ MeV,
which is too broad to be comparable with the observation. Moreover, the decay modes are inconsistent
with the observations, for example the $KK^*(1410)$ and $K^*K^*$ decay modes,
as the main modes of $\phi(2D)$, were not observed by the recent BESIII experiment~\cite{Ablikim:2020pgw}.

On the other hand, if the mass of $\phi(2170)$ is in the range of $\sim2079-2135$ MeV,
as an assignment of $\phi(2D)$, the width of $\phi(2170)$ is
predicted to be $\sim 175-225$ MeV, which is close to the upper limit of the observations.
The $KK_1(1400)$, $KK(1460)$, $KK_1(1270)$ decay modes,
as the main decay modes, were observed by the recent BESIII experiments
as well~\cite{Ablikim:2020pgw}. However, it is difficult to understand
why the other main decay mode $K^*K^*$ was not observed by the recent BESIII experiments~\cite{Ablikim:2020pgw}.
A slight mixing with the $\phi(3S)$ state (for example with a mixing angle $\theta\simeq-20^{\circ}$)
can strongly suppress partial widths of the $KK^*$, $KK^*(1410)$, and $KK^*_2(1430)$ mode, but the partial width
of $K^*K^*$ is still large, and insensitive to the mixing angle (see Fig.~\ref{widthsd}).
Confirmation of the $K^*K^*$ decay mode and accurate measurements of the resonance parameters
for $\phi(2170)$ are crucial for understanding its nature.

%The decay rates into $\phi\eta $ and $\phi\eta'$ are tiny, the branching fractions are $O(10^{-4})$.

Finally, it is should be mentioned that according to the quark model predictions
the mass difference between $\phi(2D)$ and $\phi(3S)$ states
is $\sim 100$ MeV, and both of them have a fairy broad width $\sim 200-300$ MeV.
Thus, both $\phi(2D)$ and $\phi(3S)$ may highly overlap with each other around the
energy range $\sim 2.2\pm 0.1$ GeV. Furthermore,
both $\phi(2D)$ and $\phi(3S)$ have similar strong decay properties, for example both of them
dominantly decay into $KK_1(1400)$, $KK(1460)$, $KK_1(1270)$, $KK_1(1400)$, and $KK^*_2(1430)$.
The above facts indicate that it may be hard to distinguish these two overlapping resonances from
the invariant mass distributions of the final states.
Thus, the $\phi(2170)$ resonance observed in some reactions may be a structure caused by two largely overlapping
resonances $\phi(2D)$ and $\phi(3S)$. This may explain the fairly large uncertainties for
the resonance parameters extracted from different experiments.

\subsection{$3P$-wave states}

\subsubsection{ The $3^3P_0$ $s\bar{s}$ state}

The $3^3P_0$ $s\bar{s}$ state is not established. Our quark model predicted mass is $M=2434$ MeV,
which is in good agreement with the predictions with the relativized quark model
~\cite{Godfrey:1985xj,Xiao:2019qhl}, however, is about $(80-150)$ MeV lager than
those predicted in Refs.~\cite{Ishida:1986vn,Ebert:2009ub}.
According to our analysis of the strong decay properties (see Table~\ref{decay3p2}),
this state might be a broad state with a width of
\begin{equation}\label{h1}
\Gamma_{total}\simeq 346 \ \mathrm{MeV},
\end{equation}
and dominantly decays into the $K^*K^*$, $K^*K_2^*(1430)$, $KK_1(1270)$, and $KK_2(1770)$ final states
with branching fractions $\sim12\%$, $\sim9\%$, $\sim25\%$, and $\sim27\%$, respectively.
Few study of the strong decays of the higher $3P$-wave $s\bar{s}$ states is found in the literature.

Recently, by an amplitude analysis of the $K_SK_S$ system produced in radiative $J/\psi$ decays
an isoscalar scalar $0^{++}$ state, $f_0(2410)$, with a $35\sigma$ significance was found at BESIII~\cite{Ablikim:2018izx}.
The mass and width were determined to be $M=(2411\pm17)$ MeV and
$\Gamma=(348\pm18$$^{+23}_{-1})$ MeV, respectively, which are obviously different
from those for the $f_0(2330)$ resonance listed by the PDG~\cite{Tanabashi:2018oca}.
It is interesting to find that both the mass and width of the newly observed state $f_0(2410)$ are in good agreement with
our predictions by considering it as the $3^3P_0$ $s\bar{s}$ state. Furthermore,
with this assignment we obtain sizeable branching fractions for the $KK$ and $\phi\phi$ channels, i.e.,
\begin{eqnarray}\label{h1200}
Br[f_0(2410)\to KK/\phi\phi]\simeq \mathcal{O}(1\%-2\%),
\end{eqnarray}
which may explain why the $f_0(2410)$ can be established in the $K_SK_S$ final state.
It should be mentioned that a flavor mixing between $s\bar{s}$ and
$n\bar{n}$ may exist in the $^3P_0$ state~\cite{Ricken:2003ua}, which may affect our predictions.
To better understand the nature of the $f_0(2410)$ and confirm our assignment,
more observations of some interesting decay channels, such as $\phi\phi$, $KK_1(1400)$ and $K^*K^*$,
are needed in future experiments.

\subsubsection{ The $3^3P_2$ $s\bar{s}$ state}

The $3^3P_2$ $s\bar{s}$ state is not established. Our quark model predicted mass is $M=2466$ MeV,
which is in good agreement with the predictions with the relativized quark model
~\cite{Godfrey:1985xj,Xiao:2019qhl}. The mass splitting between the $3^3P_2$ and $3^3P_0$ states is predicted to be
about $32$ MeV, which is comparable with those predicted in Refs.~\cite{Ebert:2009ub,Xiao:2019qhl}.
If the $f_0(2410)$ corresponds to the $3^3P_0$ $s\bar{s}$ state indeed, the mass of the $3^3P_2$ state might be around $2440-2450$ MeV.

According to our analysis of the strong decay properties (see Table~\ref{decay3p2}),
this state might have a moderate width of
\begin{equation}\label{h1}
\Gamma_{total}\simeq 145 \ \mathrm{MeV},
\end{equation}
and dominantly decay into the $K^*K(1460)$, $K^*K^*$, $K^*K_2^*(1430)$, $KK_3^*(1780)$,and $K^*K_1(1270)$ final states,
with branching fractions $\sim12\%$, $\sim10\%$, $\sim13\%$, $\sim11\%$, and $\sim15\%$, respectively.
The decay rates into the $KK$ and $\phi\phi$ final states might be sizeable, the branching
fractions are predicted to be about $2\%$ and $1\%$, respectively.

Some signals of the $3^3P_2$ $s\bar{s}$ state might have been
observed in some processes, such as $J/\psi\to \gamma \phi\phi/\gamma\eta\eta$ carried
out at BESIII~\cite{Ablikim:2013hq,Ablikim:2016hlu}, however,
the resonance parameters may be hard to extract from the data due to strong effects from some
nearby resonances and backgrounds.

\subsubsection{ The $3^3P_1$ $s\bar{s}$ state}

The $3^3P_1$ $s\bar{s}$ state is not established. Our quark model predicted mass is $M=2470$ MeV,
which is in good agreement with the other predictions in Refs.~\cite{Ishida:1986vn,Xiao:2019qhl}.
The mass splitting between the $3^3P_1$ and $3^3P_0$ states is predicted to be
about $30-40$ MeV in most quark models (See Table~\ref{mass}).
If the $f_0(2410)$ corresponds to the $3^3P_0$ state indeed, the mass of $3^3P_1$ might be around $2440-2450$ MeV,
it may highly overlap with the $3^3P_2$ state.
According to our analysis of the strong decay properties (see Table~\ref{decay3p1}),
this state might be a broad state with a width of
\begin{equation}\label{h1}
\Gamma_{total}\simeq 298 \ \mathrm{MeV}.
\end{equation}
The $3^3P_1$ $s\bar{s}$ state has relatively large decay rates into the
$KK^*(1410)$, $KK_2^*(1430)$, $K^*K_2^*(1430)$, $KK_2(1770)$, $KK_3^*(1780)$ channels
with a comparable branching fraction $\sim 10\%$ (details are listed in Table~\ref{decay3p1}).
The $K^*K_1(1270)$ mode may play a crucial role in the decays,
the branching fraction for this channel may reach up to $\sim 15\%$.
Moreover, the decay rates into the $KK^*$ and $\phi\phi$ final states might be sizeable, the branching
fractions are predicted to be $\sim 4.1\%$ and $\sim 1.1\%$, respectively. The
$\phi\phi$, $KK^*$, $K^*K^*$ and $KK_2^*(1430)$ channels might be good channels for
looking for the missing $3^3P_1$ $s\bar{s}$ state.

\subsubsection{ The $3^1P_1$ $s\bar{s}$ state}

The $3^1P_1$ $s\bar{s}$ state is not established. Our quark model predicted mass is $M=2435$ MeV,
which is comparable with the predictions in Refs.~\cite{Ebert:2009ub,Ishida:1986vn,Xiao:2019qhl}.
The mass splitting between the $3^1P_1$ and $3^3P_0$ states is predicted to be about a few MeV.
If the $f_0(2410)$ corresponds to the $3^3P_0$ indeed, the mass of the $3^1P_1$ state might be around $2410$ MeV as well.
According to our analysis of the strong decay properties (see Table~\ref{decay3p1}),
this state might be a broad state with a width of
\begin{equation}\label{h1}
\Gamma_{total}\simeq 269 \ \mathrm{MeV}.
\end{equation}
The $3^1P_1$ $s\bar{s}$ state has relatively large decay rates into the
$K^*K^*$, $K^*K(1460)$, $KK_2^*(1430)$, $K^*K_2(1430)$, $K^*K_1(1270)$, $K^*K_1(1400)$, $KK^*(1410)$, and $KK_3^*(1780)$ channels
with a comparable branching fraction $\sim 7-20\%$ (details are listed in Table~\ref{decay3p1}).
The decay rates into the $\phi\eta$ and $\phi\eta'$ final states are sizeable, the branching
fractions are predicted to be $\sim 1.7\%$ and $\sim 0.6\%$, respectively.
The $\phi\eta$ and $\phi\eta'$ might be good channels for our searching for the missing $3^1P_1$ $s\bar{s}$ state.

\subsection{$4S$-wave states}

\subsubsection{$4^1S_0$}

%The situation for the isoscalar $0^{-+}$ states is very complex.
The flavor mixing between $n\bar{n}$ and $s\bar{s}$ plays an important role for the low-lying isoscalar $0^{-+}$ states,
however, the spectroscopic mixing for the higher $s\bar{s}$ excitation, $4^1S_0$,
may be small~\cite{Godfrey:1985xj}. The mass for the higher $4^1S_0$ $s\bar{s}$ ($J^{PC}=0^{-+}$) state
is estimated to be $\sim 2580$ MeV within our nonrelativistic potential model, which is consistent with
the prediction with the relativized quark model~\cite{Godfrey:1985xj,Xiao:2019qhl}, but
our predicted mass are notably ($\sim150-320$ MeV) larger than that predicted in Refs.~\cite{Ishida:1986vn,Ebert:2009ub}.

Using the mass and wave function obtained from our potential model calculations,
we further estimate the strong decay properties. Our results are listed in Table~\ref{decay4s}.
It is found that the $4^1S_0$ $s\bar{s}$ has a rather broad width of
\begin{equation}\label{2d1}
\Gamma_{total}\simeq 409 \ \mathrm{MeV},
\end{equation}
and dominantly decays into the $K^*K_1(1400)$ final state with a branching fraction $\sim 15\%$.
Furthermore, the $4^1S_0$ $s\bar{s}$ state has large decay rates into the $KK^*$, $KK^*(1410)$, $K^*K_1(1270)$, $K^*K^*(1410)$, $KK^*_3(1780)$,$K^*K^*$, and $KK^*_2(1430)$
final states with comparable branching fractions of $\sim8-11\%$.

In 2016, the BESIII Collaboration observed a new resonance $X(2500)$ with a mass of
$2470^{+15}_{-19}$$^{+101}_{-23}$ MeV and a width of $230^{+64}_{-35}$$^{+56}_{-33}$ MeV
in $J/\psi\to \gamma \phi\phi$~\cite{Ablikim:2016hlu}. The preferred spin-parity numbers for the $X(2500)$
are $J^{PC}=0^{-+}$~\cite{Ablikim:2016hlu}. The newly observed state $X(2500)$ may be identified as
the $4^1S_0$  $s\bar{s}$ state. With this assignment, its measured mass, width, and
spin-parity numbers can be naturally understood in our calculations. The decay rate into the
$\phi\phi$ final state is also sizeable, the branching fraction is predicted to be
\begin{equation}\label{2d1}
Br[X(2500)\to \phi\phi]\simeq 0.71\%,
\end{equation}
which can explain why $X(2500)$ was seen in the $\phi\phi$ final state as well.

Some other interpretations of $X(2500)$, such as the fourth radial excitation of
$\eta'$ meson (i.e. $\eta'(5S)$)~\cite{Wang:2017iai}, the $5^1S_0$ $s\bar{s}$ state
~\cite{Pan:2016bac,Xue:2018jvi}, the $ss\bar{s}\bar{s}$ tetraquark state with
$J^{PC}=0^{-+}$~\cite{Lu:2019ira,Dong:2020okt}, can be found in the literature.
In some works~\cite{Li:2008we}, the $4^1S_0$ $s\bar{s}$ state was suggested to be an assignment for
the resonance $\eta(2225)$ listed by the PDG~\cite{Tanabashi:2018oca}.
To clarify the nature of $X(2500)$ and establish the $4^1S_0$ $s\bar{s}$ state, more observations of the dominant decay
modes, such as $KK^*$ and $KK^*_2(1430)$, are needed in future experiments.

\subsubsection{$4^3S_1$}

As a higher excitation, the mass of the $4^3S_1$ $s\bar{s}$ state predicted in theory spans in a large range
$\sim2.47-2.63$ GeV for less constraints from experiments. With a linear potential, both
our nonrelativistic potential model and the relativized quark model~\cite{Godfrey:1985xj,Xiao:2019qhl} give a similar mass $M\simeq 2625$ MeV.
In Ref.\cite{Ishida:1986vn}, by using a covariant oscillator quark model with one-gluon-exchange effects the authors
predicted a moderate mass $M\simeq 2540$ MeV. While, with a QCD-motivated relativistic quark model in Ref.~\cite{Ebert:2009ub}, the authors obtain a small mass $M\simeq 2472$ MeV, which is comparable with the prediction with the modified relativized quark model by replacing
the linear potential with a screening potential~\cite{Pang:2019ttv}. The observations of the $4^3S_1$ $s\bar{s}$ state
are crucial for testing the various models and developing QCD-motivated potential models.

To provide useful information for looking for the $4^3S_1$ $s\bar{s}$ state
in experiments, we further estimate the strong decay properties with the mass and wave function obtained
from our potential model calculations. Our results are listed in Table~\ref{decay4s}.
This state might have a moderate width $\Gamma\sim 227$ MeV, and mainly decays into
$K^*(892)K^*(892)$, $KK^*(892)$, $KK^*(1410)$, $K^*(892)K(1460)$,  $K^*(892)K_1(1270)$, and $K_1(1270)K_1(1270)$.
The $4^3S_1$ $s\bar{s}$ state might be found at BESIII by scanning the Born cross sections of
$e^+e^-\to K^*(892)K^*(892)$, $K^*(892)K_2^*(1430)$, $K^*(892)K_1(1270)$, $K_1(1270)K_1(1270)$
in the center-of-mass energy range $\sim2.4-2.7$ GeV.

\subsection{$2F$-wave states}

In present work, the masses of the $2F$-wave $s\bar{s}$ states are predicted to be in the range
$\sim 2525\pm 25$ MeV, the mass splitting between two different $2F$-wave states is
no more than 50 MeV. The masses for these $2F$-wave $s\bar{s}$ states were also calculated
with the relativized quark model~\cite{Xiao:2019qhl} and relativistic
quark model~\cite{Ebert:2009ub}. For a comparison, our results together
with those from Refs.~\cite{Ebert:2009ub,Xiao:2019qhl} are listed in Table~\ref{mass}.
It is found that our predicted mass splittings are in good agreement with the
predictions with relativized quark model, but our predicted masses are
about $80$ MeV larger than those predicted in Ref.~\cite{Xiao:2019qhl}. Although our predicted masses
for the $2^1F_3$, $2^3F_2$, and $2^3F_3$ states are close to those predicted in Refs.~\cite{Ebert:2009ub},
the predicted mass splittings are very different.

There are few discussions in respect to decay properties of the $2F$-wave states in the literature.
To provide useful information for looking for the $2F$-wave $s\bar{s}$ states
in experiments, we further estimate the strong decay properties with the mass and wave function obtained
from our potential model calculations. Our results are listed in Tables~\ref{decay2fa} and~\ref{decayfb}.
In the $2F$-wave $s\bar{s}$ states, the $2^3F_4$ has a relatively narrow width of $\Gamma\sim 145$ MeV,
the $2^3F_2$ has a very broad width of $\Gamma\sim 490$ MeV,
while the other two $2F$-wave states $2^1F_3$, and $2^3F_3$ have a comparable decay width of
$\Gamma\sim 250$ MeV.

Many OZI-allowed two-body strong decay channels are open for these $2F$-wave states.
The $2^3F_4$ state may dominantly decay into the $KK(1460)$, $KK^*(1410)$, $K^*K^*$,
$K^*K_2^*(1430)$, $KK_1(1270)$, $K^*K_1(1270)$, $K^*K_1(1400)$, $KK_2(1820)$ and $KK^*_3(1780)$ channels with comparable branching fractions
$\sim 6-13\%$ (details seen in Table~\ref{decay2fa}). The decays of $2^3F_2$ are governed by the
$K_1(1270)K_1(1270)$ channel with a large branching fraction $\sim 45\%$. The $2^3F_2$ also
has sizeable decay rates into $K^*K_2^*(1430)$, $K^*K_1(1270)$, and  $KK_2(1770)$ with branching
fractions $\sim 4\%$, $\sim 9\%$, and $\sim 17\%$, respectively.
The $2^3F_3$ state may dominantly decay into the $KK^*(1410)$, $K^*K(1460)$,
$K^*K_2^*(1430)$, $K^*K_1(1270)$, and $KK_3^*(1780)$ channels with a comparable branching fraction
$\sim 7-15\%$. The $2^1F_3$ state mostly decays into the $KK^*(1410)$, $KK^*_2(1430)$, $K^*K_1(1270)$ and $KK_3^*(1780)$ channels with a comparable branching fraction $\sim 13-20\%$, and also has fairly large decay rates into the $KK^*$, $K^*K(1460)$, $K^*K_2^*(1430)$, and $K^*K_1(1400)$ channels with a comparable branching fraction $\sim 6\%$.
It should be pointed out that our predictions for these high mass excitations may be strongly model dependent because
there are no constrains from the experiments.

\subsection{$3D$-wave states}

The $3D$-wave $s\bar{s}$ states in the present investigation are predicted to be
largely overlapping states with a mass round $\sim 2.7$ GeV, our predicted masses
are comparable with those predicted with the relativized quark model~\cite{Xiao:2019qhl} and relativistic
quark model~\cite{Ebert:2009ub}. Our predicted mass splitting between any two $3D$-wave states is no more than 20 MeV,
which is consistent with GI model~\cite{Xiao:2019qhl}, while it is smaller than the
predictions in Ref~\cite{Ebert:2009ub}.

There are few discussions in respect to the decay properties of the $3D$-wave states in the literature.
To provide useful information for looking for the $3D$-wave $s\bar{s}$ states
in experiments, we further estimate the strong decay properties with the mass and wave function obtained
from our potential model calculations. Our results are listed in Tables~\ref{decay12}, \ref{decay3da}, \ref{decay3db}, and~\ref{decay3dc}.
In the $3D$-wave $s\bar{s}$ states, both $3^3D_3$ and $3^1D_2$ have a comparable decay width of $\Gamma\sim 160-200$ MeV,
the $3^3D_2$ state has a relatively broad width of $\Gamma\sim 270$ MeV, while
the $3^3D_1$ state might a rather broad state with a width of $\Gamma\sim 350$ MeV.

Many OZI-allowed two-body strong decay channels are open for these $3D$-wave states.
The $3^3D_1$ state may dominantly decay into $K^*K^*(1410)$, $K^*K_2^*(1430)$, $KK_1(1270)$, $K^*K_1(1270)$ and $KK_2(1770)$
with branching fractions $\sim 18\%$, $\sim 10\%$, $\sim 8\%$, $\sim 15\%$,
and $\sim 17\%$, respectively. These dominant decay modes and their decay rates for $3^3D_1$
are notably different from those predicted in Ref.~\cite{Pang:2019ttv} due to different resonance mass
adopted in the calculations. The $3^3D_2$ state may dominantly decay into the $K^*K^*(1410)$, $K^*K_2^*(1430)$, $K^*K_1(1270)$, $K_1(1270)K_1(1270)$,
$KK_3^*(1780)$, and $K^*K_2(1770)$ final states with branching fractions $\sim 12\%$, $\sim 8\%$, $\sim 11\%$, $\sim 13\%$, $\sim 11\%$,
and $\sim 6\%$, respectively. The $3^3D_3$ state may dominantly decay into the $KK(1460)$, $K^*K^*(1410)$, $K^*K_2^*(1430)$, $KK_1(1270)$,
and $K^*K_3^*(1780)$ final states with a comparable branching fraction $\sim 8-13\%$.
The $3^1D_2$ state may dominantly decay into the $K^*K^*(1410)$, $KK_2^*(1430)$, $K^*K_2^*(1430)$, $K^*K_1(1270)$,
and $KK_3^*(1780)$ final states with branching fractions $\sim 13\%$, $\sim 8\%$, $\sim 8\%$, $\sim 13\%$,
and $\sim 16\%$, respectively. It should be pointed out that our predictions for these high mass excitations
may be strongly model dependent because there are no constrains from the experiments.

%\subsection{$P$-wave states}

\section{summary}\label{sum}

In this paper we calculate the $s\bar{s}$ spectrum up to the mass range
of $\sim 2.7$ GeV with a nonrelativistic linear quark potential model, where the
model parameters are partially adopted from a calculation of the $\Omega$ spectrum.
Then, with the widely used $^3P_0$ model we further analyze the OZI-allowed two-body strong decays of the
$s\bar{s}$ states by using wave functions obtained from the potential model.
Based on our successful explanations of the well established states $\phi(1020)$, $\phi(1680)$, $h_1(1415)$,
$f'_2(1525)$, and $\phi_3(1850)$, we further discuss the possible assignments of
strangeonium-like states from experiments by combining our theoretical results
with the observations. We expect that our present study can deepen
our knowledge about the $s\bar{s}$ spectrum and provides useful references for looking for the missing $s\bar{s}$ in future experiments.
Several key points of this work are emphasized as follows:

\begin{itemize}

\item Some isoscalar $0^{++}$ resonances with mass of $\sim 1370$ MeV
(denoted with $f_0(1370)$ by the PDG) observed in the $KK$ and $\eta\eta$ final states may
correspond to the $1^3P_0$ $s\bar{s}$ state.

\item The $f_2(2010)$ listed by the PDG~\cite{Tanabashi:2018oca} might be a good candidate for the $2^3P_2$
$s\bar{s}$ state. The newly observed $1^{+-}$ resonance $X(2062)$ in the $\eta' \phi$
mass spectrum of the decay $J/\psi\to \phi \eta \eta'$ at BESIII~\cite{Ablikim:2018xuz}
favors the assignment of the $2^1P_1$ $s\bar{s}$ state.

\item The isoscalar scalar $0^{++}$ state with a mass of $M=(2411\pm17)$ MeV
(denoted with $f_0(2410)$) observed in $J/\psi\to K_SK_S$ at BESIII~\cite{Ablikim:2018izx} may be
a newly observed state different from the $f_0(2330)$ resonance listed by the
PDG~\cite{Tanabashi:2018oca}. The $f_0(2410)$ favors the assignment of the $3^3P_0$ $s\bar{s}$ state.

\item The broad resonance $f_2(2150)$ listed by the PDG~\cite{Tanabashi:2018oca} can be assigned as the
$1^3F_2$ $s\bar{s}$ state. Another relatively narrow $4^{++}$ resonance $f_4(2210)$ first observed in
the reaction $K^-p\to K^+K^- \Lambda$ by the LASS Collaboration~\cite{Aston:1988yp} might be
an assignment of the $1^3F_4$ $s\bar{s}$ state.

\item  The new resonance $X(2500)$ observed in $J/\psi\to \gamma \phi\phi$ at BESIII~\cite{Ablikim:2016hlu}
may be identified as the $4^1S_0$ $s\bar{s}$ state.

\item  The possibility of $\phi(2170)$ as a candidate for $\phi(3S)$ or $\phi(2D)$ cannot be excluded.
Further observations of the $K^*K^*$ decay mode, and precise measurements of the resonance parameters
and branching ratios between the main decay modes for the $\phi(2170)$ state are crucial to confirm its nature.

\end{itemize}

\begin{table*}[htb]
\caption{ Strong decay properties for the  $1S$-, $2S$-, $3S$-wave vector $s\bar s$ states. $\Gamma_{th}$ and $Br$ stand for the partial widths and branching fractions of the strong decay processes, respectively. The experimental widths $\Gamma_{exp}$ are taken from the PDG~\cite{Tanabashi:2018oca}. To know the contributions of different partial waves to a decay process, the partial wave amplitudes of every decay mode (denoted with Amps.) are also given in the table. For a comparison, some other predictions within the $~^{3}P_{0}$~~Model~\cite{Barnes:2002mu,Pang:2019ttv} are also listed.}\label{decay1}
\begin{tabular}{cccccccccccccccccccccc}   \midrule[1.0pt]\midrule[1.0pt]
 {\multirow{2}{*}{State}} & \multirow{2}{*}{Mode}~~~~& \multicolumn{3}{c} {\underline{~~~~~~~~~~~~~~~~~~~~~~~~~~~~~~Ours~~~~~~~~~~~~~~~~~~~~~~~~~~~~}}
 & \multicolumn{2}{c} {\underline{~~~~~~~~~~~~Ref.~\cite{Barnes:2002mu}~~~~~~~~~~~~}}
 & \multicolumn{2}{c} {\underline{~~~~~~~~~~~~Ref.~\cite{Pang:2019ttv}~~~~~~~~~~~~}}&
 \\
    &
    &$\Gamma_{th}$ ~[$\Gamma_{exp}$] (MeV)  &$Br(\%)$ &Amps. (GeV$^{-1/2}$) &$\Gamma_{th}$ (MeV)&	$Br(\%)$&$\Gamma_{th}$ (MeV)&$Br(\%)$\\
\midrule[1.0pt]	 % meson    % mode		            	       	     %Gamma(ours)  %Br(ours)     %Gamma TB     %Br  TB	    %Gamma Pang	  %Br pang
$1 ^3S_{1}[1020]$
                                    &   $KK$                     ~~~~&   4.09~[3.5]      ~~~~&  100         ~~~~&$^1P_1$=-0.079 ~~~~&2.5             ~~~~&59        ~~~~&$\cdots$ ~~~~&$\cdots$ \\
%                                   &   Total                    ~~~~&   $\cdots$  ~~~~&  $\cdots$    ~~~~&               ~~~~&$\cdots$            ~~~~&$\cdots$        ~~~~&$\cdots$ ~~~~&$\cdots$ \\
\\
$2 ^3S_{1}$$[1680]$                 &   $KK$                     ~~~~&   8.06  ~~~~&  5      ~~~~&$^1P_1$=0.062  ~~~~&  89            ~~~~&23.54      ~~~~&  15.5   ~~~~&    10.33 \\
                                    &   $KK^*(892)$              ~~~~&   132   ~~~~&  79     ~~~~&$^3P_1$=-0.215 ~~~~&  245           ~~~~&64.81      ~~~~&  117    ~~~~&    78    \\
                                    &   $\phi\eta$               ~~~~&   26.3  ~~~~&  16     ~~~~&$^3P_1$=-0.537 ~~~~&  44            ~~~~&11.64      ~~~~&  16.7   ~~~~&    11.13 \\
                                    &   Total                    ~~~~&   167 ~[$150\pm50$]  ~~~~&  100    ~~~~&               ~~~~&  378           ~~~~&100        ~~~~&  150    ~~~~&    100   \\
\end{tabular}
\end{table*}

\begin{table*}[htb]
\caption{Strong decay properties for the $1P$ and $2P$-wave $s\bar s$ states.}\label{decay3}
\begin{tabular}{cccccccccccccccccccccc}  \midrule[1.0pt]\midrule[1.0pt]
Mode & State & $\Gamma_{th}$~[$\Gamma_{exp}$] (MeV) & $Br(\%)$ & Amps. (GeV$^{-1/2}$) &  State & $\Gamma_{th}$~[$\Gamma_{exp}$](MeV) & $Br(\%)$ & Amps. (GeV$^{-1/2}$)
 \\
\midrule[1.0pt]		
~~~~~~$KK$                  &$1 ^3P_{0}$  ~~~~~~ &281     ~~~~~~~& 83     ~~~~~~~&$^1S_0$=-0.392 ~~~~~~&$1 ^3P_{2}$   ~~~~~& 40.3   ~~~~~~& 70      ~~~~~~&$^1D_2$=0.142 \\
~~~~~~$KK^*(892)$           &(1373)       ~~~~~~ &$\cdots$~~~~~~~&$\cdots$~~~~~~~&               ~~~~~~&(1525)        ~~~~~& 12.2   ~~~~~~& 21      ~~~~~~&$^3D_2$=-0.076\\
~~~~~~$\eta\eta$            &             ~~~~~~ &56.2    ~~~~~~~& 17     ~~~~~~~&$^1S_0$=0.332  ~~~~~~&              ~~~~~& 5.45   ~~~~~~& 9       ~~~~~~&$^1D_2$=-0.096\\
~~~~~~$\eta\eta'$           &             ~~~~~~ &$\cdots$~~~~~~~&$\cdots$~~~~~~~&               ~~~~~~&              ~~~~~& 0.01   ~~~~~~& 0.02    ~~~~~~&$^1D_2$=-0.011\\
~~~~~~Total                 &             ~~~~~~ &338 [200-500]    ~~~~~~~&100     ~~~~~~~&               ~~~~~~&              ~~~~~& 58~[$86\pm5$]   ~~~~~~& 100     ~~~~~~&              \\
\\
~~~~~~$KK^*(892)$           &$1 ^3P_{1}$  ~~~~~~ &381     ~~~~~~~& 100    ~~~~~~~&$^3S_1$=-0.453 ~~~~~~&$1 ^1P_{1}$   ~~~~~& 141    ~~~~~~& 100     ~~~~~~&$^3S_1$=-0.361\\
\end{tabular}
\end{table*}

\begin{table*}[htb]
\caption{Strong decay properties for the $1D$-wave $s\bar s$ states. For a comparison, some other predictions within $~^{3}P_{0}$~~Model~\cite{Barnes:2002mu,Pang:2019ttv} are also listed.}\label{decay61}
 \begin{tabular}{cccccccccccccccccccccc}   \midrule[1.0pt]\midrule[1.0pt]
 {\multirow{2}{*}{State}} & \multirow{2}{*}{Mode}~~~~& \multicolumn{3}{c} {\underline{~~~~~~~~~~~~~~~~~~~~~~~~~~~~~~Ours~~~~~~~~~~~~~~~~~~~~~~~~~~~~}}
 & \multicolumn{2}{c} {\underline{~~~~~~~~~~~~Ref.~\cite{Barnes:2002mu}~~~~~~~~~~~~}}
 & \multicolumn{2}{c} {\underline{~~~~~~~~~~~~Ref.~\cite{Pang:2019ttv}~~~~~~~~~~~~}}&
 \\
 &&$\Gamma_{th}$~[$\Gamma_{exp}$] (MeV)  &$Br(\%)$ &Amps. (GeV$^{-1/2}$) &$\Gamma_{th}$ (MeV)&	$B_{r}(\%)$&$\Gamma_{th}$ (MeV)&$Br(\%)$\\
\midrule[1.0pt]	 % meson    % mode		            	    %Gamma(ours)  %Br(ours)     %Gamma TB     %Br  TB	    %Gamma Pang	  %Br pang
$1 ^3D_{1}$(1809)$\footnote{Mass (MeV) adopted in present work.}$
                                    &   $KK$                    ~~~~&   30.5  ~~~~&  4     ~~~~&$^1P_1$=-0.117~~~~&  65     ~~~~&9.97     ~~~~&  40.8   ~~~~&7.46                  \\
($1850\footnote{Mass (MeV) adopted in Ref.~\cite{Barnes:2002mu}.}/1869\footnote{Mass (MeV) adopted in Ref.~\cite{Pang:2019ttv}.}$)
                                    &   $KK^*(892)$             ~~~~&   42.0  ~~~~&  6     ~~~~&$^3P_1$=-0.113~~~~&  75     ~~~~&11.50    ~~~~&  57.8   ~~~~&10.57                 \\
                                    &   $K^*(892)K^*(892)$      ~~~~&   1.02  ~~~~&  0.1   ~~~~&$^1P_1$=-0.048~~~~&  5      ~~~~&0.77     ~~~~&  11.5   ~~~~&2.10                  \\
                                    &                           ~~~~&         ~~~~&        ~~~~&$^5P_1$=0.022 ~~~~&         ~~~~&         ~~~~&         ~~~~&                      \\
                                    &                           ~~~~&         ~~~~&        ~~~~&$^5F_1$=0.003 ~~~~&         ~~~~&         ~~~~&         ~~~~&                      \\
                                    &   $\phi\eta$              ~~~~&   13.2  ~~~~&  2     ~~~~&$^3P_1$=-0.322~~~~&  29     ~~~~&4.45     ~~~~&  13.6   ~~~~&2.49                  \\
\end{tabular}
\end{table*}

\begin{table*}[htb]
\caption{Strong decay properties for the $1F$-wave states.}\label{decay1f2}
\label{decay12}
\begin{center}
\begin{tabular}{ccccccccccccccccc}\midrule[1.0pt]\midrule[1.0pt]
Mode & State & $\Gamma_{th}$ (MeV) & $Br(\%)$ & Amps. (GeV$^{-1/2}$) &  State & $\Gamma_{th}$ (MeV) & $Br(\%)$ & Amps. (GeV$^{-1/2}$) \\
\midrule[1.0pt]
~~~~~~$KK$              &$1 ^3F_{2}$  ~~~~~~&   5.40  ~~~~~~& 1.8     ~~~~~~&$^1D_2$=0.048 ~~~~~~&$1 ^3F_{4}$   ~~~~~~&   7.8   ~~~~~~& 11      ~~~~~~~~~&$^1G_4$=-0.058\\
~~~~~~$KK^*(892)$       &  (2146)     ~~~~~~&   12.2  ~~~~~~& 4       ~~~~~~&$^3D_2$=0.056 ~~~~~~&  (2078)      ~~~~~~&   16    ~~~~~~& 23      ~~~~~~~~~&$^3G_4$=0.064 \\
~~~~~~$KK(1460)$        &             ~~~~~~&   9.37  ~~~~~~& 3       ~~~~~~&$^1D_2$=-0.049~~~~~~&              ~~~~~~&   0.13  ~~~~~~& 0.2     ~~~~~~~~~&$^1G_4$=0.006 \\
~~~~~~$KK^*(1410)$      &             ~~~~~~&   0.57  ~~~~~~& 0.2     ~~~~~~&$^3D_2$=-0.016~~~~~~&              ~~~~~~& $<0.01$ ~~~~~~& $<0.01$ ~~~~~~~~~&$^3G_4\sim$0  \\
~~~~~~$K^*(892)K^*(892)$&             ~~~~~~&   19.4  ~~~~~~& 6.3     ~~~~~~&$^1D_2$=0.054 ~~~~~~&              ~~~~~~&   33    ~~~~~~& 47      ~~~~~~~~~&$^1G_4$=-0.017\\
~~~~~~                  &             ~~~~~~&         ~~~~~~&         ~~~~~~&$^5D_2$=-0.041~~~~~~&              ~~~~~~&         ~~~~~~&         ~~~~~~~~~&$^5D_4$=0.152 \\
~~~~~~                  &             ~~~~~~&         ~~~~~~&         ~~~~~~&$^5G_2$=-0.094~~~~~~&              ~~~~~~&         ~~~~~~&         ~~~~~~~~~&$^5G_4$=0.034 \\
\end{tabular}
\end{center}
\end{table*}

\begin{table*}[htb]
\caption{Strong decay properties for the $2 ^3D_{1}$ and $2 ^3D_{3}$ states.}\label{decay2d1}
\begin{tabular}{cccccccccccccccccccccc}  \midrule[1.0pt]\midrule[1.0pt]
Mode & State & $\Gamma_{th}$ (MeV) & $Br(\%)$ & Amps. (GeV$^{-1/2}$) &  State & $\Gamma_{th}$ (MeV) & $Br(\%)$ & Amps. (GeV$^{-1/2}$) \\
\midrule[1.0pt]	        % meson		    % mode		            			%Gamma		     	    %Br		        %   mass		
~~~~~~$KK$               & $2 ^3D_{1}$ ~~~~~~&   7.8  ~~~~~~&  2.4    ~~~~~~&$^1P_1$=-0.057~~~~~~&  $2 ^3D_{3}$ ~~~~~~& 7.7     ~~~~~~&  5.7    ~~~~~~&$^1F_3$=0.057 \\
~~~~~~$KK^*(892)$        & (2272)      ~~~~~~&   7.7  ~~~~~~&  2.4    ~~~~~~&$^3P_1$=-0.043~~~~~~&    (2285)    ~~~~~~& 3.6     ~~~~~~&  2.7    ~~~~~~&$^3F_3$=-0.030\\
~~~~~~$KK(1460)$         &             ~~~~~~&   38   ~~~~~~&  12     ~~~~~~&$^1P_1$=-0.089~~~~~~&              ~~~~~~& 25      ~~~~~~&  19     ~~~~~~&$^1F_3$=0.072 \\
~~~~~~$KK^*(1410)$       &             ~~~~~~&   32   ~~~~~~&  9.8    ~~~~~~&$^3P_1$=-0.092~~~~~~&              ~~~~~~& 18      ~~~~~~&  13     ~~~~~~&$^3F_3$=-0.067\\
~~~~~~$K^*(892)K^*(892)$ &             ~~~~~~&   28   ~~~~~~&  8.6    ~~~~~~&$^1P_1$=-0.023~~~~~~&              ~~~~~~& 17      ~~~~~~&  12     ~~~~~~&$^1F_3$=-0.014\\
~~~~~~                   &             ~~~~~~&        ~~~~~~&         ~~~~~~&$^5P_1$=0.010 ~~~~~~&              ~~~~~~&         ~~~~~~&         ~~~~~~&$^5P_3$=-0.095\\
~~~~~~                   &             ~~~~~~&        ~~~~~~&         ~~~~~~&$^5F_1$=-0.128~~~~~~&              ~~~~~~&         ~~~~~~&         ~~~~~~&$^5F_3$=0.032 \\
\end{tabular}
\end{table*}

\begin{table*}[htb]
\caption{Strong decay properties for the $2 ^3D_{2}$ and $2 ^1D_{2}$ states.}\label{decay11}
\begin{tabular}{cccccccccccccccccccccc}  \midrule[1.0pt]\midrule[1.0pt]
Mode & State & $\Gamma_{th}$ (MeV) & $Br(\%)$ & Amps. (GeV$^{-1/2}$) &  State & $\Gamma_{th}$ (MeV) & $Br(\%)$ & Amps. (GeV$^{-1/2}$) \\
\midrule[1.0pt]	        % meson		    % mode		            			%Gamma		     	    %Br		        %   mass		                                    \\
~~~~~~$KK^*(892)$        & $2 ^3D_{2}$ ~~~~~~& 15      ~~~~~~& 6.5     ~~~~~~&$^3P_2$=-0.058~~~~~~&  $2 ^1D_{2}$ ~~~~~~&  11      ~~~~~~& 5.2     ~~~~~~~~~&$^3P_2$=-0.048\\
~~~~~~                   &  (2297)     ~~~~~~&         ~~~~~~&         ~~~~~~&$^3F_2$=0.016 ~~~~~~&    (2282)    ~~~~~~&          ~~~~~~&         ~~~~~~~~~&$^3F_2$=-0.017\\
~~~~~~$KK^*(1410)$       &             ~~~~~~& 71      ~~~~~~& 30      ~~~~~~&$^3P_2$=-0.118~~~~~~&              ~~~~~~&  57      ~~~~~~& 27      ~~~~~~~~~&$^3P_2$=-0.100\\
~~~~~~                   &             ~~~~~~&         ~~~~~~&         ~~~~~~&$^3F_2$=0.061 ~~~~~~&              ~~~~~~&          ~~~~~~&         ~~~~~~~~~&$^3F_2$=-0.069\\
~~~~~~$K^*(892)K^*(892)$ &             ~~~~~~& 11      ~~~~~~& 4.6     ~~~~~~&$^5P_2$=0.044 ~~~~~~&              ~~~~~~&  12      ~~~~~~& 5.6     ~~~~~~~~~&$^3P_2$= 0.060\\
~~~~~~                   &             ~~~~~~&         ~~~~~~&         ~~~~~~&$^5F_2$=-0.067~~~~~~&              ~~~~~~&          ~~~~~~&         ~~~~~~~~~&$^3F_2$=0.060 \\
~~~~~~$KK^*_0(1430)$     &             ~~~~~~& $<0.01$ ~~~~~~& $<0.01$ ~~~~~~&$^1D_2\sim$0  ~~~~~~&              ~~~~~~&  9.2     ~~~~~~& 4.4     ~~~~~~~~~&$^1D_2$=0.045 \\
\end{tabular}
\end{table*}

\begin{table*}[!htbp]
\caption{Partial and total strong decay widths (MeV) for the $\phi(2170)$ as candidates for $3 ^3S_{1}$ and $2 ^3D_{1}$, respectively. Case I, Case II, Case III stand for the results by taking the mass of $\phi(2170)$ with 2079 MeV, 2135 MeV, and 2175 MeV respectively. The values in the bracket
are branching fractions.}   \label{decay2170}
\begin{tabular}{cccccccccccccccccccccccccc}  \midrule[1.0pt]\midrule[1.0pt]
\multirow{2}{*}{Mode}
~~&\multicolumn{2}{c} {\underline{~~~~~~~~~~~~~~Case I~~(2079 MeV)~~~~~~~~~~~~~~~}}
~~&\multicolumn{2}{c} {\underline{~~~~~~~~~~~~~~Case II~~(2135 MeV)~~~~~~~~~~~~~~}}
~~&\multicolumn{2}{c} {\underline{~~~~~~~~~~~~~~Case III~~(2175 MeV)~~~~~~~~~~~~~~}}
\\
~&
 $3 ^3S_{1}$ ~&$2 ^3D_{1}$
~&$3 ^3S_{1}$ ~&$2 ^3D_{1}$
~&$3 ^3S_{1}$ ~&$2 ^3D_{1}$
\\
\midrule[1.0pt]
$KK$                   & 2.90   ~(1.8\%)  ~~~~~ & 9.9   ~(6.3\%)  ~~~~~ &2.4   ~(1\%)     ~~~~~ & 9.6   ~(4.5\%)   ~~~~~ &1.9   ~(0.7\%)   ~~~~~ & 9.2   ~(3.2\%)  \\
$KK^*(892)$            & 27     ~(17\%)   ~~~~~ & 5.2   ~(3.3 \%) ~~~~~ &29    ~(12\%)    ~~~~~ & 6.3   ~(3\%)     ~~~~~ &30    ~(11 \%)   ~~~~~ & 6.9    ~(2.4 \%)  \\
$KK(1460)$             & 35     ~(22\%)   ~~~~~ & 59    ~(38\%)   ~~~~~ &35    ~(15\%)    ~~~~~ & 66    ~(31\%)    ~~~~~ &29    ~(10 \%)   ~~~~~ & 62    ~(22\%)   \\
$KK^*(1410)$           & 2.1    ~(1.4\%)  ~~~~~ & 0.58  ~(0.37\%) ~~~~~ &62    ~(26\%)    ~~~~~ & 17    ~(7.9\%)   ~~~~~ &96    ~(35\%)    ~~~~~ & 27    ~(9\%)   \\
$K^*(892)K^*(892)$     & 2.9    ~(1.8\%)  ~~~~~ & 14    ~(9.2\%)  ~~~~~ &11    ~(4.9\%)   ~~~~~ & 20    ~(9.2\%)   ~~~~~ &21    ~(7.5\%)   ~~~~~ & 23    ~(8\%)  \\
$KK^*_2(1430)$         & 28     ~(18\%)   ~~~~~ & 11    ~(7.3\%)  ~~~~~ &42    ~(18\%)    ~~~~~ & 18    ~(8.5 \%)  ~~~~~ &49    ~(18 \%)   ~~~~~ & 22    ~(8\%)   \\
\end{tabular}
\end{table*}

\begin{table*}[htb]
\caption{Strong decay properties for the $3 ^3P_{0}$ and $3 ^3P_{2}$ $s\bar s$ states.}\label{decay3p2}
\begin{tabular}{cccccccccccccccccccccc}  \midrule[1.0pt]\midrule[1.0pt]
Mode & State & $\Gamma_{th}$ (MeV) & $Br(\%)$ & Amps. (GeV$^{-1/2}$) &  State & $\Gamma_{th}$ (MeV) & $Br(\%)$ & Amps. (GeV$^{-1/2}$)
 \\
\midrule[1.0pt]		
~~~~~~$KK$                  &$3 ^3P_{0}$  ~~~~~~ &9.2     ~~~~~~~&2.7     ~~~~~~~&$^1S_0$=-0.062 ~~~~~~&$3 ^3P_{2}$   ~~~~~ & 3.0    ~~~~~~&2.1     ~~~~~~&$^1D_2$=0.035  \\
~~~~~~$KK^*(892)$           &(2434)       ~~~~~~ &$\cdots$~~~~~~~&$\cdots$~~~~~~~&$\cdots$       ~~~~~~& (2466)       ~~~~~ & 0.1    ~~~~~~&0.1     ~~~~~~&$^3D_2$=0.005 \\
~~~~~~$KK(1460)$            &             ~~~~~~ &30      ~~~~~~~&8.6     ~~~~~~~&$^1S_0$=-0.072 ~~~~~~&              ~~~~~ & 3.7    ~~~~~~&2.6     ~~~~~~&$^1D_2$=0.025  \\
~~~~~~$KK^*(1410)$          &             ~~~~~~ &$\cdots$~~~~~~~&$\cdots$~~~~~~~&$\cdots$       ~~~~~~&              ~~~~~ & 3.2    ~~~~~~&2.2     ~~~~~~&$^3D_2$=0.025  \\
~~~~~~$K^*(892)K^*(892)$    &             ~~~~~~ &39      ~~~~~~~&12      ~~~~~~~&$^1S_0$=-0.025 ~~~~~~&              ~~~~~ & 15     ~~~~~~&10      ~~~~~~&$^1D_2$=-0.021 \\
~~~~~~                      &             ~~~~~~ &        ~~~~~~~&        ~~~~~~~&$^5D_0$=-0.145 ~~~~~~&              ~~~~~ &        ~~~~~~&        ~~~~~~&$^5S_2$=-0.068 \\
~~~~~~                      &             ~~~~~~ &        ~~~~~~~&        ~~~~~~~&               ~~~~~~&              ~~~~~ &        ~~~~~~&        ~~~~~~&$^5D_2$=0.056  \\
\end{tabular}
\end{table*}

\begin{table*}[htb]
\caption{Strong decay properties for the $3 ^3P_{1}$ and $3 ^1P_{1}$ $s\bar s$ states.}\label{decay3p1}
\begin{tabular}{cccccccccccccccccccccc}  \midrule[1.0pt]\midrule[1.0pt]
Mode & State & $\Gamma_{th}$ (MeV) & $Br(\%)$ & Amps. (GeV$^{-1/2}$) &  State & $\Gamma_{th}$ (MeV) & $Br(\%)$ & Amps. (GeV$^{-1/2}$)
 \\
\midrule[1.0pt]		
~~~~~~$KK^*(892)$           &$3 ^3P_{1}$  ~~~~~~ & 12     ~~~~~~~& 4.1    ~~~~~~~&$^3S_1$=-0.051 ~~~~~~&$3 ^1P_{1}$   ~~~~~ &  12    ~~~~~~& 4.6    ~~~~~~&$^3S_1$=-0.038 \\
~~~~~~                      &             ~~~~~~ &        ~~~~~~~&        ~~~~~~~&$^3D_1$=-0.016 ~~~~~~&              ~~~~~ &        ~~~~~~&        ~~~~~~&$^3D_1$=0.037  \\
~~~~~~$KK^*(1410)$          &(2470)       ~~~~~~ & 31     ~~~~~~~& 10     ~~~~~~~&$^3S_1$=-0.074 ~~~~~~& (2435)       ~~~~~ &  28    ~~~~~~& 10     ~~~~~~&$^3S_1$=-0.050 \\
~~~~~~                      &             ~~~~~~ &        ~~~~~~~&        ~~~~~~~&$^3D_1$=-0.025 ~~~~~~&              ~~~~~ &        ~~~~~~&        ~~~~~~&$^3D_1$=0.056  \\
~~~~~~$K^*(892)K^*(892)$    &             ~~~~~~ & 20     ~~~~~~~& 6.7    ~~~~~~~&$^5D_1$=-0.105 ~~~~~~&              ~~~~~ &  18    ~~~~~~& 6.6    ~~~~~~&$^3S_1$=-0.040 \\
~~~~~~                      &             ~~~~~~ &        ~~~~~~~&        ~~~~~~~&               ~~~~~~&              ~~~~~ &        ~~~~~~&        ~~~~~~&$^3D_1$=0.092  \\
~~~~~~$K^*(892)K(1460)$     &             ~~~~~~ & 12     ~~~~~~~& 4.0    ~~~~~~~&$^3S_1$=0.013  ~~~~~~&              ~~~~~ &  18    ~~~~~~& 6.6    ~~~~~~&$^3S_1$=0.032  \\
\end{tabular}
\end{table*}

\begin{table*}[htb]
\caption{Strong decay properties for the $4S$-wave $s\bar s$ states.}\label{decay4s}
\begin{tabular}{cccccccccccccccccccccc}  \midrule[1.0pt]\midrule[1.0pt]
Mode & State & $\Gamma_{th}$ (MeV) & $Br(\%)$ & Amps. (GeV$^{-1/2}$) &  State & $\Gamma_{th}$ (MeV) & $Br(\%)$ & Amps. (GeV$^{-1/2}$)
 \\
\midrule[1.0pt]	
~~~~~~$KK$                  & $4 ^1S_{0}$  ~~~~~~ &$\cdots$~~~~~~~&$\cdots$~~~~~~~&$\cdots$       ~~~~~~&$4 ^3S_{1}$  ~~~~~ & 0.42   ~~~~~~&  0.18   ~~~~~~&$^1P_1$=0.013 \\
~~~~~~$KK^*(892)$           & (2580)       ~~~~~~ & 42     ~~~~~~~& 10     ~~~~~~~&$^3P_0$=-0.098 ~~~~~~&  (2623)$^a$ ~~~~~ & 12     ~~~~~~&  5.5    ~~~~~~&$^3P_1$=-0.053\\
~~~~~~$KK(1460)$            &              ~~~~~~ &$\cdots$~~~~~~~&$\cdots$~~~~~~~&$\cdots$       ~~~~~~&             ~~~~~ & 0.49   ~~~~~~&  0.22   ~~~~~~&$^1P_1$=0.009 \\
~~~~~~$KK^*(1410)$          &              ~~~~~~ & 46     ~~~~~~~& 11     ~~~~~~~&$^3P_0$=-0.090 ~~~~~~&             ~~~~~ & 19     ~~~~~~&  8.3    ~~~~~~&$^3P_1$=-0.057\\
~~~~~~$K^*(892)K^*(892)$    &              ~~~~~~ & 30     ~~~~~~~& 7.3    ~~~~~~~&$^3P_0$=-0.126 ~~~~~~&             ~~~~~ & 29     ~~~~~~&  13     ~~~~~~&$^1P_1$=0.027 \\
~~~~~~                      &              ~~~~~~ &        ~~~~~~~&        ~~~~~~~&               ~~~~~~&             ~~~~~ &        ~~~~~~&         ~~~~~~&$^5P_1$=-0.119\\
~~~~~~$K^*(892)K(1460)$     &              ~~~~~~ & 6.4    ~~~~~~~& 1.6    ~~~~~~~&$^3P_0$=-0.039 ~~~~~~&             ~~~~~ & 11     ~~~~~~&  4.9    ~~~~~~&$^3P_1$=0.050 \\
\end{tabular}
\end{table*}

\begin{table*}[htb]
\caption{Strong decay properties for the $3 ^3D_{3}$ state.}\label{decay12}
\begin{center}
\begin{tabular}{ccccccccccccccccc}\midrule[1.0pt]\midrule[1.0pt]
 State &Mode & $\Gamma_{th}$ (MeV) & $Br(\%)$ & Amps. (GeV$^{-1/2}$) &Mode  & $\Gamma_{th}$ (MeV) & $Br(\%)$ & Amps. (GeV$^{-1/2}$) \\
\midrule[1.0pt]
 $3 ^3D_{3}$ (2691) &$KK$                  &  6.1   &3.8     &$^1F_3$=0.050  &$KK^*(1680)$          & 0.04     & 0.02    &$^3F_3$=-0.003  \\
 (2691)             &$KK^*(892)$           &  3.8   &2.4     &$^3F_3$=-0.029 &$KK_2(1770)$          & 6.4      & 4.0     &$^5P_3\sim$=0   \\
                    &$KK(1460)$            &  13    &8.1     &$^1F_3$=0.044  &                      &          &         &$^5F_3$=-0.003  \\
                    &$KK^*(1410)$          &  6.8   &4.3     &$^3F_3$=-0.033 &                      &          &         &$^5H_3$=-0.034  \\
                    &$K^*(892)K^*(892)$    &  7.2   &4.5     &$^1F_3$= 0.004 &$KK_2(1820)$          & 8.4      & 5.3     &$^5P_3$=-0.038  \\
                    &                      &        &        &$^5P_3$=-0.059 &                      &          &         &$^5F_3$=0.013   \\
                    &                      &        &        &$^5F_3$= 0.010 &                      &          &         &$^5H_3$=0.003   \\
\end{tabular}
\end{center}
\end{table*}

\begin{table*}[htb]
\caption{Strong decay properties for the $3 ^3D_{1}$ state.}\label{decay3da}
\begin{center}
\begin{tabular}{ccccccccccccccccc}\midrule[1.0pt]\midrule[1.0pt]
 State &Mode & $\Gamma_{th}$ (MeV) & $Br(\%)$ & Amps. (GeV$^{-1/2}$) &Mode  & $\Gamma_{th}$ (MeV) & $Br(\%)$ & Amps. (GeV$^{-1/2}$) \\
\midrule[1.0pt]
$3 ^3D_{1}$(2681)  & $KK$                  &   2.8   &  0.82  &$^1P_1$=-0.034  & $K_1(1270)K_1(1400)$  &   3.0   &  0.87  &$^1P_1$=0.046  \\
                   & $KK^*(892)$           &   3.3   &  0.94  &$^3P_1$=-0.027  &                       &         &        &$^3P_1$=0.024  \\
                   & $KK(1460)$            &   4.1   &  1.2   &$^1P_1$=-0.025  &                       &         &        &$^5P_1$=0.011  \\
                   & $KK^*(1410)$          &   4.6   &  1.3   &$^3P_1$=-0.027  &                       &         &        &$^5F_1\sim$0   \\
                   & $K^*(892)K^*(892)$    &   16    &  4.7   &$^1P_1$=-0.021  & $KK^*(1680)$          &   0.08  &  0.02  &$^3P_1$=-0.004 \\
                   &                       &         &        &$^5P_1$=0.009   & $KK_2(1770)$          &   59    &  17    &$^5P_1$=-0.102 \\
                   &                       &         &        &$^5F_1$=-0.088  &                       &         &        &$^5F_1$=0.018  \\
\end{tabular}
\end{center}
\end{table*}

\begin{table*}[htb]
\caption{Strong decay properties for the $3 ^3D_{2}$ state.}\label{decay3db}
\begin{center}
\begin{tabular}{ccccccccccccccccc}\midrule[1.0pt]\midrule[1.0pt]
 State &Mode & $\Gamma_{th}$ (MeV) & $Br(\%)$ & Amps. (GeV$^{-1/2}$) &Mode  & $\Gamma_{th}$ (MeV) & $Br(\%)$ & Amps. (GeV$^{-1/2}$) \\
\midrule[1.0pt]
$3 ^3D_{2}$ (2701)  &$KK^*(892)$           & 5.2     &  1.9    &$^3P_2$=-0.031 & $KK^*(1680)$         & 3.1     &  1.2     &$^3P_2$=0.021  \\
                    &                      &         &         &$^3F_2$=0.013  &                      &         &          &$^3F_2$=0.009  \\
                    &$KK^*(1410)$          & 14      &  5.3    &$^3P_2$=-0.040 & $KK_2(1770)$         & 7.6     &  2.9     &$^5P_2$=0.034  \\
                    &                      &         &         &$^3F_2$=0.027  &                      &         &          &$^5F_2$=0.014  \\
                    &$K^*(892)K^*(892)$    & 4.6     &  1.7    &$^5P_2$=0.031  & $KK_2(1820)$         & 1.4     &  0.52    &$^5P_2$=0.002  \\
                    &                      &         &         &$^5F_2$=-0.037 &                      &         &          &$^5F_2$=-0.017 \\
                    &$K^*(892)K(1460)$     & 12      &  4.4    &$^3P_2$=0.047  & $KK^*_3(1780)$       & 30      &  11      &$^7P_2$=-0.066 \\
\end{tabular}
\end{center}
\end{table*}

\begin{table*}[htb]
\caption{Strong decay properties for the $3 ^1D_{2}$ state.}\label{decay3dc}
\begin{center}
\begin{tabular}{ccccccccccccccccc}\midrule[1.0pt]\midrule[1.0pt]
 State &Mode & $\Gamma_{th}$(MeV) & $Br(\%)$ & Amps. (GeV$^{-1/2}$) &Mode  & $\Gamma_{th}$(MeV) & $Br(\%)$ & Amps. (GeV$^{-1/2}$) \\
\midrule[1.0pt]
$3 ^1D_{2}$(2685)   &$KK^*(892)$           &  4.0    & 2       &$^3P_2$=-0.026 &$KK^*(1680)$          &   7.1   & 3.6      &$^3P_2$=-0.032 \\
                    &                      &         &         &$^3F_2$=-0.014 &                      &         &          &$^3F_2$=0.013  \\
                    &$KK^*(1410)$          &   12    & 6       &$^3P_2$=-0.035 &$KK_2(1770)$          &   0.24  & 0.12     &$^5P_2$=-0.007 \\
                    &                      &         &         &$^3F_2$=-0.028 &                      &         &          &$^5F_2\sim$0   \\
                    &$K^*(892)K^*(892)$    &   5.8   & 2.9     &$^3P_2$=-0.042 &$KK_2(1820)$          &   0.25  & 0.13     &$^5P_2$=-0.007 \\
                    &                      &         &         &$^3F_2$=0.034  &                      &         &          &$^5F_2$=-0.002 \\
                    &$K^*(892)K(1460)$     &   9.0   & 4.5     &$^3P_2$=-0.038 &$KK^*_3(1780)$        &   32    & 16       &$^7P_2$=-0.063 \\
\end{tabular}
\end{center}
\end{table*}

\begin{table*}[htb]
\caption{Strong decay properties for the $2 ^3F_{2}$ and $2 ^3F_{4}$ states.}\label{decay2fa}
\begin{center}
\begin{tabular}{ccccccccccccccccc}\midrule[1.0pt]\midrule[1.0pt]
Mode & State & $\Gamma_{th}$ (MeV) & $Br(\%)$ & Amps. (GeV$^{-1/2}$) &  State & $\Gamma_{th}$ (MeV) & $Br(\%)$ & Amps. (GeV$^{-1/2}$) \\
\midrule[1.0pt]
~~~~~~$KK$                 &$2 ^3F_{2}$  ~~~~~~&   2.3   ~~~~~~&  0.46   ~~~~~~&$^1D_2$=0.031 ~~~~~~&$2 ^3F_{4}$   ~~~~~~&   5.3   ~~~~~~&  3.6    ~~~~~~~~~&$^1G_4$=-0.047\\
~~~~~~$KK^*(892)$          &  (2552)     ~~~~~~&   3.9   ~~~~~~&  0.79   ~~~~~~&$^3D_2$=0.030 ~~~~~~&  (2503)      ~~~~~~&   6.4   ~~~~~~&  4.4    ~~~~~~~~~&$^3G_4$=0.038 \\
~~~~~~$KK(1460)$           &             ~~~~~~&   1.3   ~~~~~~&  0.27   ~~~~~~&$^1D_2$=0.015 ~~~~~~&              ~~~~~~&   13    ~~~~~~&  8.7    ~~~~~~~~~&$^1G_4$=-0.046\\
~~~~~~$KK^*(1410)$         &             ~~~~~~&   6.0   ~~~~~~&  1.2    ~~~~~~&$^3D_2$=0.033 ~~~~~~&              ~~~~~~&   19    ~~~~~~&  13     ~~~~~~~~~&$^3G_4$=0.061 \\
~~~~~~$K^*(892)K^*(892)$   &             ~~~~~~&   1.9   ~~~~~~&  0.38   ~~~~~~&$^1D_2$=0.024 ~~~~~~&              ~~~~~~&   9.9   ~~~~~~&  6.8    ~~~~~~~~~&$^1G_4$=-0.002\\
~~~~~~                     &             ~~~~~~&         ~~~~~~&         ~~~~~~&$^5D_2$=-0.018~~~~~~&              ~~~~~~&         ~~~~~~&         ~~~~~~~~~&$^5D_4$=0.073 \\
~~~~~~                     &             ~~~~~~&         ~~~~~~&         ~~~~~~&$^5G_2$= 0.009~~~~~~&              ~~~~~~&         ~~~~~~&         ~~~~~~~~~&$^5G_4$=0.004 \\
\end{tabular}
\end{center}
\end{table*}

\begin{table*}[htb]
\caption{Strong decay properties for the $2 ^3F_{3}$ and $2 ^1F_{3}$ states.}\label{decayfb}
\begin{center}
\begin{tabular}{ccccccccccccccccc}\midrule[1.0pt]\midrule[1.0pt]
 State &Mode & $\Gamma_{th}$ (MeV) & $Br(\%)$ & Amps. (GeV$^{-1/2}$) &Mode  & $\Gamma_{th}$ (MeV) & $Br(\%)$ & Amps. (GeV$^{-1/2}$) \\
\midrule[1.0pt]
$2 ^3F_{3}$ (2543)  &$KK^*(892)$           &   11    &  4      &$^3D_3$=0.035  & $K^*(892)K_1(1400)$  &   13     &  4.9    &$^1F_3$=-0.009  \\
                    &                      &         &         &$^3G_3$=-0.035 &                      &          &         &$^3F_3$=0.054   \\
                    &$KK^*(1410)$          &   28    &  10     &$^3D_3$=0.039  &                      &          &         &$^5P_3$=-0.007  \\
                    &                      &         &         &$^3G_3$=-0.060 &                      &          &         &$^5F_3$=-0.017  \\
                    &$K^*(892)K^*(892)$    &   3.4   &  1.3    &$^5D_3$=-0.042 &                      &          &         &$^5H_3$=-0.003  \\
                    &                      &         &         &$^5G_3$=-0.007 & $KK^*(1680)$         &   2.8    &  1.1    &$^3D_3$=-0.023  \\
                    &$K^*(892)K(1460)$     &   20    &  7.4    &$^3D_3$=-0.064 &                      &          &         &$^3G_3$=-0.005  \\
\end{tabular}
\end{center}
\end{table*}

\section*{Acknowledgements}

The authors thank Dr. Wen-Biao Yan and Long-Sheng Lu for very helpful discussions.
This work is supported by the National Natural Science Foundation of China under Grants
No.~U1832173, No.~11775078, No.~11705056, and No.~11405053.

\end{document}